\documentclass[12pt]{article}

\usepackage{scicite}
\usepackage{url}

\usepackage{times}

\usepackage{amsmath}
\usepackage{amsfonts}
\usepackage{amssymb}
\usepackage{graphicx}
\usepackage{subcaption}
\usepackage{xcolor}

\definecolor{p1}{HTML}{66C2A5} 
\definecolor{p2}{HTML}{FC8D62} 
\definecolor{p3}{HTML}{8DA0CB} 

\newenvironment{sciabstract}{
\begin{quote} \bf}
{\end{quote}}

\title{Well-Connected Communities in Real-World and Synthetic Networks}

\author
{Minhyuk Park,${}^{1\textsuperscript{\textdagger}}$ Yasamin Tabatabaee,${}^{1\textsuperscript{\textdagger}}$ Vikram Ramavarapu,${}^{1\textsuperscript{\textdagger}}$\\
Baqiao Liu,${}^{1}$ Vidya K. Pailodi,${}^{1}$ Rajiv Ramachandran,${}^{1}$ Dmitriy Korobskiy,${}^{2}$ \\
Fabio Ayres,${}^{3}$ George Chacko,${}^{1,4\ast}$ Tandy Warnow${}^{1\ast}$\\
\\
\normalsize{${}^{1}$Department of Computer Science, University of Illinois Urbana-Champaign, IL 61801, USA}\\
\normalsize{${}^{2}$NTT DATA, McLean, VA 22102, USA}\\
\normalsize{${}^{3}$Insper Institute, S\~{a}o Paulo, Brazil}\\
\normalsize{${}^{4}$Grainger College of Engineering, University of Illinois Urbana-Champaign, IL 61801, USA}\\
\\
\normalsize{$^\ast$To whom correspondence should be addressed; E-mail:  chackoge@illinois.edu; warnow@illinois.edu}\\
\normalsize{$\textsuperscript{\textdagger}$Contributed equally to this manuscript.}
}

\date{}

\begin{document} 


\baselineskip24pt


\maketitle


\newpage

\begin{sciabstract} 
Integral to the problem of detecting communities through graph clustering is the expectation that they are 
“well connected”. In this respect, we examine five different community detection approaches optimizing different criteria: the Leiden algorithm optimizing the Constant Potts Model, the Leiden algorithm optimizing modularity, Iterative K-Core Clustering (IKC), Infomap, and Markov Clustering (MCL).  
Surprisingly, all these methods produce, to varying extents, communities that fail even a mild requirement for well connectedness. To remediate clusters that are not well connected, we have developed the “Connectivity Modifier” (CM), which, at the cost of coverage,  iteratively removes small edge cuts and re-clusters until all communities produced are well connected. Results from real-world and synthetic networks illustrate a tradeoff users make between well connected clusters and coverage, and raise questions about the ``clusterability'' of networks and models of community structure. 
\end{sciabstract}

\newpage
\paragraph* {Introduction} Community detection is of broad interest and is typically posed as a graph partitioning problem, where the input is a  graph and the objective is a partitioning of its vertices into disjoint subsets, so that each subset represents a community \cite{newman2004finding,Mucha2010,Fortunato2022}. The terms community and cluster overlap heavily, so we use them interchangeably herein. Our interest in community detection is for the purpose of identifying research communities from the global scientific literature, so we are 
especially focused on methods that can scale to large networks consisting of documents linked by citation \cite{Wedell2022,Boyack2013,Waltman2012,Boyack2011}.

A unifying definition of community does not exist  but a general expectation is that  the vertices within a community are better connected to each other than to vertices outside the community \cite{Coscia2011}, implying greater edge density within a community. However,   a cluster may be dense while still having a small edge cut \cite{Bonchi2021}. 
Therefore, the minimum edge cut size (min cut) for a community should not be small \cite{traag2019louvain}. 
Thus, {\em edge density} and {\em well-connectedness},  i.e., not having a small edge cut, are two {\em separable} and expected properties of communities.

The Leiden algorithm \cite{traag2019louvain}, which builds upon the Louvain algorithm \cite{Blondel2008}, is commonly used for community detection, with default quality function the Constant Potts Model (CPM) \cite{Traag2011}.
Clusters produced by CPM-optimization have the desirable property that if the edge cut splits the cluster into components $A$ and $B$, then the edge cut will be at least $r \times |A| \times |B|$ \cite{traag2019louvain}, where $r$ is a user-provided resolution parameter.  
This guarantee is strong when the edge cut splits a cluster into two components of approximately equal size, but is weaker when it produces an imbalanced split and weakest when the cut separates a single node from the remaining nodes in the cluster. Importantly,  the guarantee depends on $r$, and small values of $r$ produce weak bounds. Finally, we note that this guarantee applies to CPM-optimal clusterings but not to clusterings found by heuristics. 

In using the Leiden software optimizing CPM, we observed that it produces clusters with small min cuts on seven different networks of varied origin ranging in size from approximately 34,000 to 75 million nodes.
We also observed that the number of clusters  with small min cuts increases as the resolution parameter decreases. Intrigued by this observation, we performed a broader study to evaluate the extent to which clusters produced by algorithms of interest meet even a mild standard for a well connected cluster. 

To evaluate whether a cluster is well connected, we use a slow growing function $f(n)$ so that a cluster with $n$ nodes whose min cut size is at most $f(n)$ will not be
considered well connected.
By design, we ensure that (i) $f(n)$ grows more slowly than the lower bound on the min cut size for clusters in CPM-optimal clusterings in the Leiden algorithm and (ii) that $f(n)$ provides a meaningful lower bound on the small-to-moderate values of $n$ where the bound in \cite{traag2019louvain} is weak. 
We selected $f(n) =  \log_{10}  n  $ for this function.

We constructed min cut profiles from four additional clustering methods with different optimality criteria on the seven networks above: Leiden with modularity \cite{newman2004finding} as quality function; the \emph{k-core} based Iterative $k$-core Clustering (IKC)  \cite{Wedell2022} using $k=10$; and two flow-based methods, Infomap \cite{Rosvall2008} and Markov clustering (MCL) \cite{VanDongen2008}. None of these methods offer guarantees of well connected clusters. 
While only IKC and Leiden optimizing either CPM or modularity  scaled to the largest network we studied, all tested methods produced poorly connected clusters (i.e., clusters with min cuts of size at most $f(n)$) on these networks. 
 \emph{These observations reveal a gap between the expectation of well connected clusters and what is actually being produced by these community finding methods.} These findings also raise questions about the ``clusterability'' \cite{Miasnikof2023} of networks and whether only portions of a network exhibit community structure. 

For practical remediation, we developed a tunable meta-method, the Connectivity Modifier (CM). CM takes a clustering as input and returns well connected clusters that are also at least  a minimum size $B$, where the setting for $B$ as well as the definition of ``well connected'' can be set by the user   (our default setting uses the function $f(n)$ given above and $B=11$).
CM presently provides support for Leiden optimizing either CPM or modularity and IKC, the methods that scaled to the largest network we studied.  On real-world networks, using CM in conjunction with the Leiden method produces well connected clusters but with a reduction in node coverage. We observed similar results of lower magnitude with IKC. 
Analyses of synthetic networks with ground truth communities show somewhat different trends, revealing intriguing differences between synthetic and real-world networks.

The rest of this manuscript is organized as follows. First, we present an initial study on a large citation network showing conditions under which Leiden clusters are not well connected. Next we present comparable results from additional methods and networks. We then describe the design of Connectivity Modifier (CM) and show the impact of using CM on clusterings by Leiden and IKC on real world and synthetic networks. Finally, we close with a discussion of our findings.

\section*{Results}

\begin{figure}[ht]
\centering
\begin{subfigure}[t]{0.45\textwidth}
\begin{center}
\includegraphics[width=0.9\linewidth]{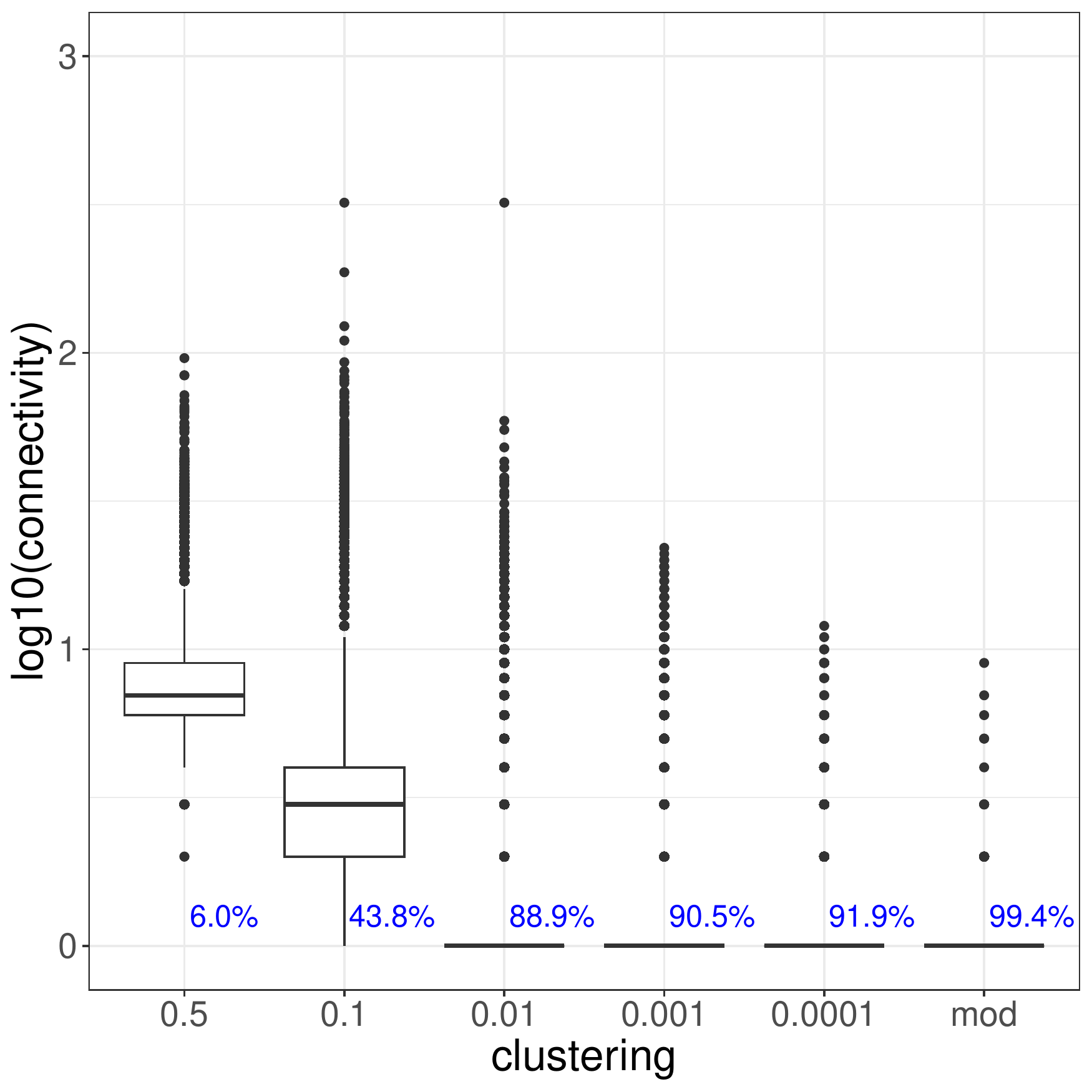}
\end{center}
\end{subfigure}
\begin{subfigure}[t]{0.45\textwidth}
\begin{center}
\includegraphics[width=0.9 \linewidth]{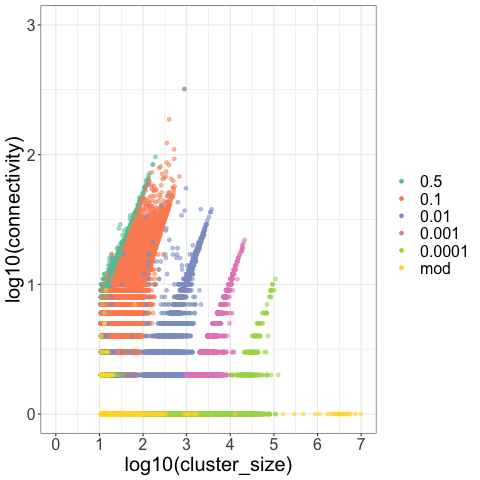}
\end{center}
\end{subfigure}
\caption{\emph{Node coverage, connectivity, and size distribution of clusters generated by Leiden optimizing either CPM or modularity on the Open Citations network (75,025,194 nodes).} Connectivity (y-axis) is the minimum edge cut size (min cut) of each cluster.  Node coverage, the percentage of nodes in clusters of at least size 11, is reported in blue text. Results are shown for clusters of at least size 11 from Leiden optimizing either CPM at five different resolution values or modularity. Higher node coverage is associated with reduced connectivity. Within each clustering, larger clusters are better connected although lower resolution values and modularity trend towards larger clusters that are less well connected. }
\label{fig:prelim-study}
\end{figure}

In a study of the Open Citations network \cite{Peroni2020} consisting of  75,025,194 nodes, we computed the min cut of all clusters generated using the Leiden algorithm optimizing either CPM or modularity (Fig.~1). 
Designating singleton clusters and very small clusters as not being of practical interest, 
we report node coverage throughout this manuscript as the percentage of nodes in clusters of size at least 11; the exception to this is Figure 4, where we  report node coverage based on non-singleton clusters. 
For CPM, we used five different resolution values ($0.5, 0.10, 0.01, 0.001,$ and $0.0001$) that resulted in node coverage values ranging from 6-99\%. 

For Leiden-CPM clusterings, we see that as the resolution value is decreased, (i) node coverage increases, (ii) the frequency of small mincuts increases, and (iii) cluster sizes increase (Fig.~1). Clustering under modularity is most similar to clustering under CPM at the lowest resolution value used. Strikingly, 98.7\% of 2,184 clusters produced under modularity had a minimum edge cut  size of 1  (i.e., could be split by removing a single edge) while accounting for 99.4\% node coverage. 
Additionally, where node coverage is high, clusters tend to be larger and fewer although intermediate resolution values in the range we used result in  an increase in the number of clusters (Fig.~1 and Table S1). 
Thus, these data illustrate a tradeoff that users make with the Leiden algorithm between small clusters, lower node coverage, and few small min cuts  (achieved by CPM-optimization with larger resolution values) versus larger clusters, higher node coverage, and many more small min cuts (achieved by modularity-optimization or CPM-optimization with small resolution values).

While a large fraction of small min cuts intuitively signals ``poorly-connected'' clusters, there are different ways of formalizing this notion. 
Here we offer a formal definition for the purposes of this study.  
Briefly, we consider functions $f(n)$ with the interpretation that if a cluster of size $n$ has an edge cut of size at most  $f(n)$ then the cluster will be considered poorly connected. We want $f(n)$ to grow very slowly so that it serves as a mild bound. We  also want  $f(n) \geq 1$ for all $n$ that are large enough for the cluster to be considered a potential community. From three examples of slowly growing functions (Materials and Methods), we choose $f(n)= \log_{10}n $, the function that imposes the mildest constraint  on large clusters and grows more slowly than the bound in \cite{traag2019louvain}. 

In addition to the Open Citations network, we clustered six other networks, ranging in size from 34,546 nodes to 13,989,436 nodes,  with Leiden, IKC, Infomap, and MCL and computed the percentage of clusters that we consider well connected (i.e., whose min cuts were greater than  $f(n)$).  
Under the conditions used, Leiden and IKC ran to completion on all seven networks, although IKC did not return any clusters from wiki\_talk because a 10-core does not exist in this sparse network. Infomap failed on the largest network, and MCL returned output only from the smallest network (cit\_hepph) we analyzed (Fig.~2). 

For Leiden clustering  optimizing CPM (Fig.~1), the frequency of well connected clusters decreases with resolution value, and results from modularity are similar to the lowest resolution value for CPM that was tested. IKC completed on all networks returning well connected clusters that varied between 85.9\% and 94\% of the total number of clusters. In comparison to Leiden, IKC clustering resulted in lower node coverage, which is consistent with its more conservative formulation. Infomap produced well connected clusters varying from 5\% (orkut) to 92.4\%  (cit\_patents). MCL ran only on the cit\_hepph network with 81.3\% of the clusters being well connected. Interestingly, while neither Leiden nor IKC generated disconnected clusters, Infomap generated disconnected clusters for some networks (maximized at 75\% for orkut) and  MCL also generated disconnected clusters (7.2\%) on the cit\_hepph network. These observations reveal the widespread existence of clusters that are not 
well connected, with the extent dependent on clustering method and network.

\begin{figure}[ht]
\centering
\begin{subfigure}[t]{0.45\textwidth}
\begin{center}
\includegraphics[width=0.95\linewidth]{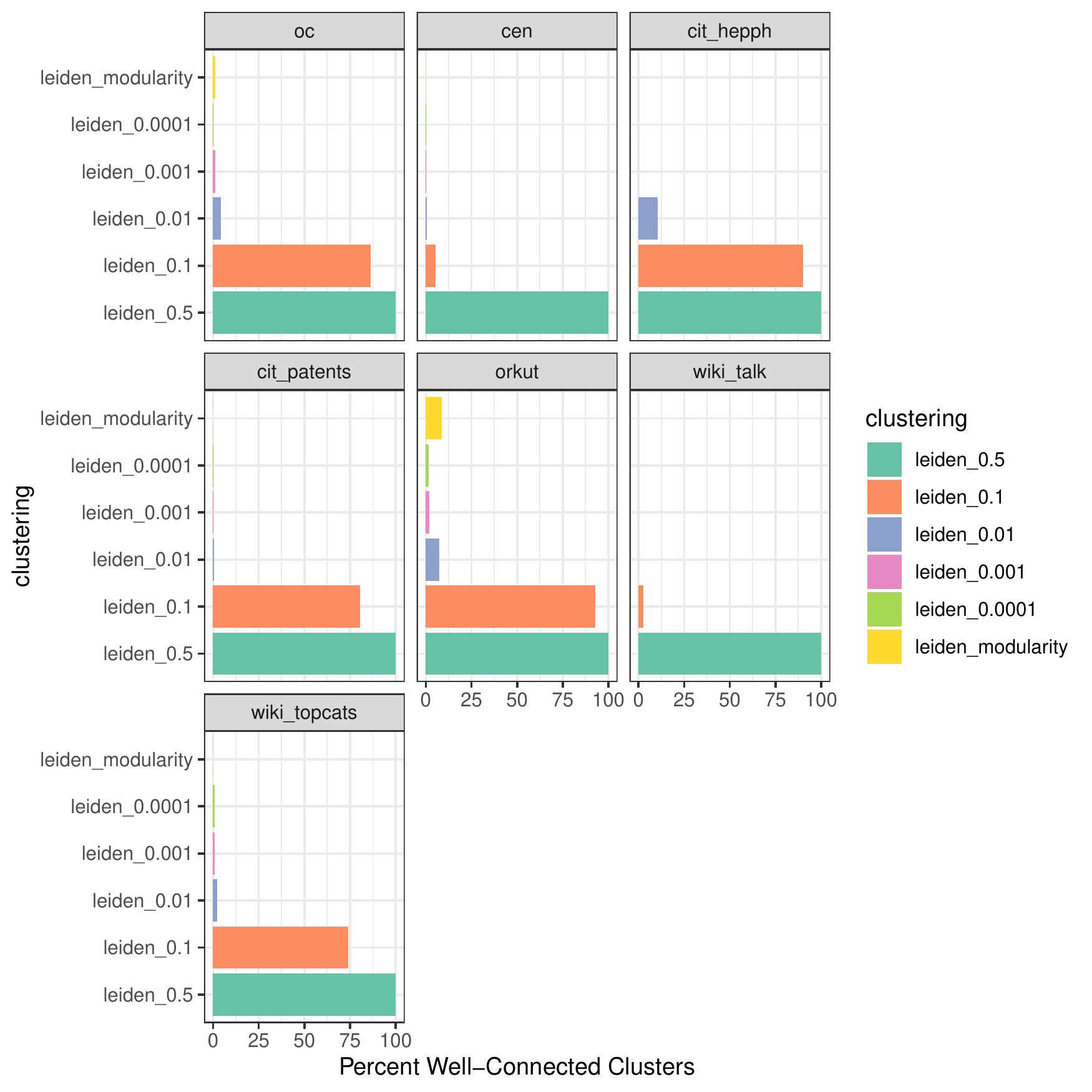}
\caption{Leiden CPM \& Modularity}
\end{center}
\end{subfigure} 
\begin{subfigure}[t]{0.45\textwidth}
\begin{center}
\includegraphics[width=0.95\linewidth]{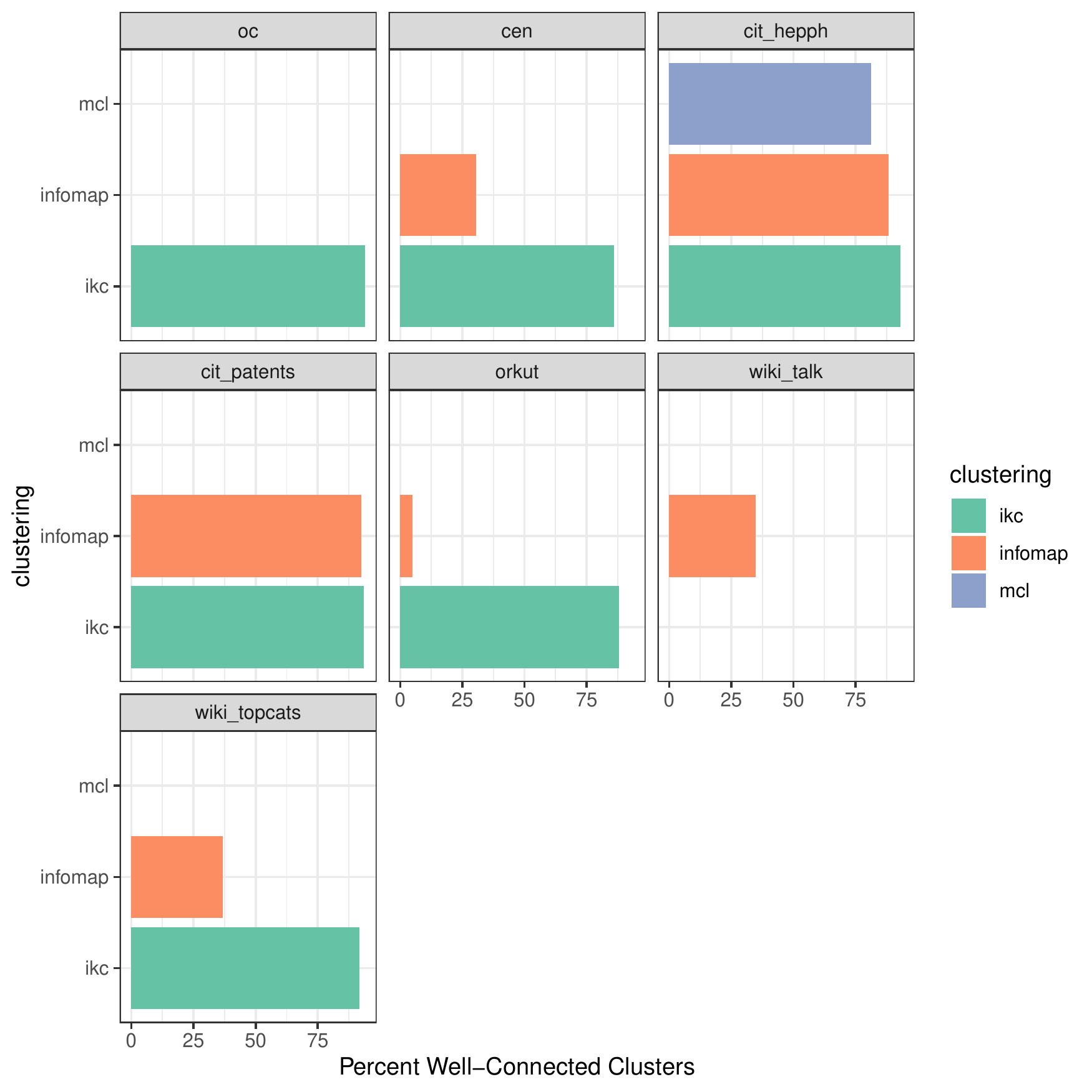}
\caption{IKC, Infomap, MCL}
\end{center}
\end{subfigure}\hspace{0.1\textwidth}
\caption{\emph{Percentage of Well Connected Clusters.} Clustering of seven networks by five different community finding approaches. Networks analyzed are  Open Citations (75,025,194), Curated Exosome Network (13,989,436), cit\_hepph (34,546), and cit\_patents (3,774,768) are citation networks; orkut (3,072,441) is a social network; wiki\_talk (2,394,385) and wiki\_topcats (1,791,489) are Wikipedia communication and hyperlink networks respectively. Only Leiden and IKC ran to completion on all networks although IKC did not return any clusters from the wiki\_talk network. Infomap completed on all but Open Citations. MCL completed only on cit\_hepph. }
\label{fig:seven-networks}
\end{figure}

Towards well connected clusters, we designed the Connectivity Modifier (CM) \cite{cm2023,cm2022} a remediation tool that can be used to modify a given clustering to ensure that each final cluster is well connected   (Materials and Methods). Based on our preliminary findings,  we chose to initially evaluate CM paired with the Leiden and IKC methods since these two clustering methods were sufficiently scalable and did not produce disconnected clusters. We also restricted our attention to the two largest networks, Open Citations and CEN.

We implemented CM in a pipeline (Fig.~3) that presently takes a Leiden or IKC clustering as input. A pre-processing (filtering) step discards very small clusters, i.e., those of size at most 10, but this bound can be changed by the user. 
Tree clusters are also discarded in this pre-processing step (given our definition of $f(n)$, any tree of size 10 or larger is  not well connected). 
CM then iteratively computes and removes any min cuts of size at most $f(n)$ and re-clusters until only well connected clusters remain. A post-processing step removes any small clusters of size at most 10 that may have resulted from repeated cutting.

\begin{figure}[ht]
\centering
\includegraphics[width=0.9\linewidth]{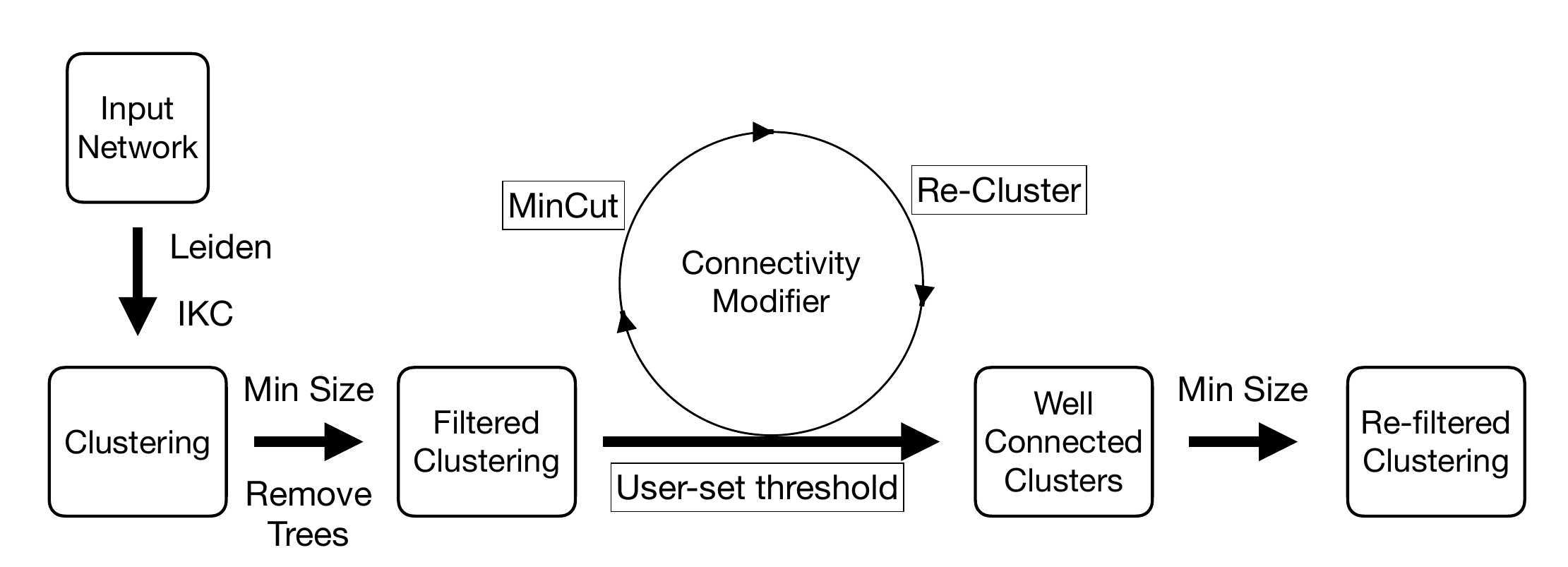}
\caption{\emph{Connectivity Modifier Pipeline Schematic.} The four-stage pipeline depends on user-specified algorithmic parameters: $B$, the minimum allowed size of a cluster, and $f(n)$, a bound on the minimum edge cut size for a cluster with $n$ nodes, and clustering method.  \emph{Stage 1}: a clustering is computed. \emph{Stage 2}: clusters are pre-processed by removing trees and those clusters of size less than $B$. \emph{Stage 3}: the CM is applied to each cluster, removing edge cuts of sizes at most $f(n)$,  reclustering, and recursing. \emph{Stage 4}: clusters are post-processed by removing those of size less than $B$. All clusters returned are well connected according to  $f(n)$ and have size at least $B$.  Our study explored default settings with $B=11$ and $f(n) =   \log_{10}n $.}
\label{fig:schematic}
\end{figure}
\clearpage
.
\paragraph{Effect of CM on node coverage} We assessed the impact of CM on node coverage, here specifically examining clusters of size at least two.  As above, we examined Leiden clustering of Open Citations and CEN networks using CPM-optimization at five resolution values and also using modularity,  examining the change in node coverage before and after CM treatment (Fig.~4).  The results are resolution dependent and network sensitive. 
For CPM-optimization, the impact of filtering out clusters of size at most 10 is large for the two larger resolution values, but then decreases as the resolution value decreases.
In contrast, the impact of the  Connectivity Modifier (the middle component of the CM pipeline that iteratively finds and removes small edge cuts and reclusters) also depends on the resolution parameter, with a minimal impact for the large resolution values and an increasing impact as the resolution value decreases.
Modularity returned results most similar to CPM-optimization with the smallest tested resolution value.
Since the pre-processing (filtering) and post-processing both remove all clusters of size at most 10, the node coverage reported for these stages are with respect to 
clusters of size at least 11.
Given this, we observe that post-CM node coverage is low compared to pre-CM for both networks and clustering methods, and was smallest when using CPM-optimization with
resolution value $r=0.5$ and largest when using CPM-optimization with one of the two smallest resolution values,  $r=0.001$ for CEN and  $r=0.0001$ for Open Citations. Overall, post-CM node coverage  of any Leiden clustering never exceeded  24.6\% for CEN and 68.7\% for Open Citations.

\begin{figure}[ht]
\centering
\begin{subfigure}[t]{0.45\textwidth}
\begin{center}
\includegraphics[width=1.0\linewidth]{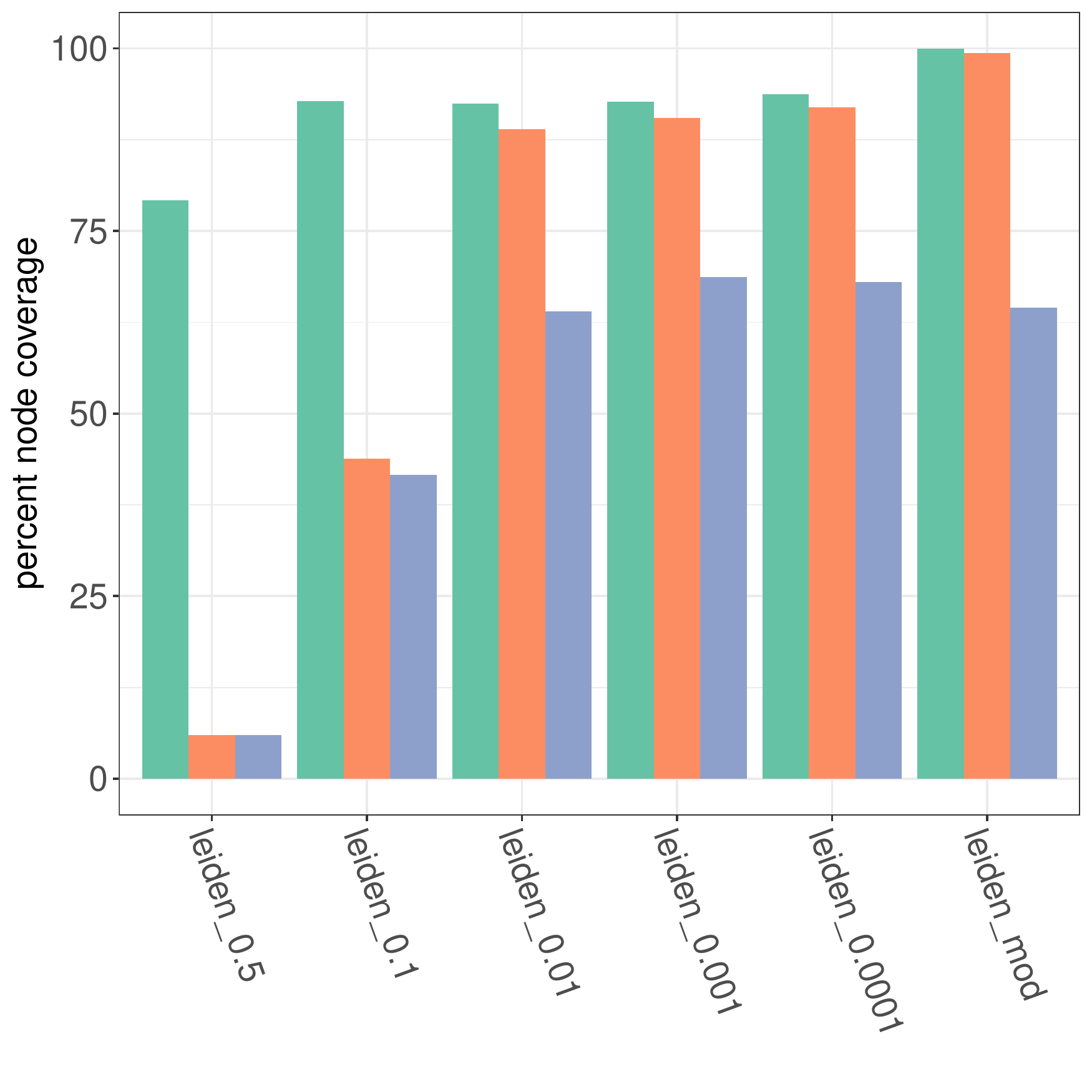}
\caption{Open Citations}
\end{center}
\end{subfigure}
\begin{subfigure}[t]{0.45\textwidth}
\begin{center}
\includegraphics[width=1.0\linewidth]{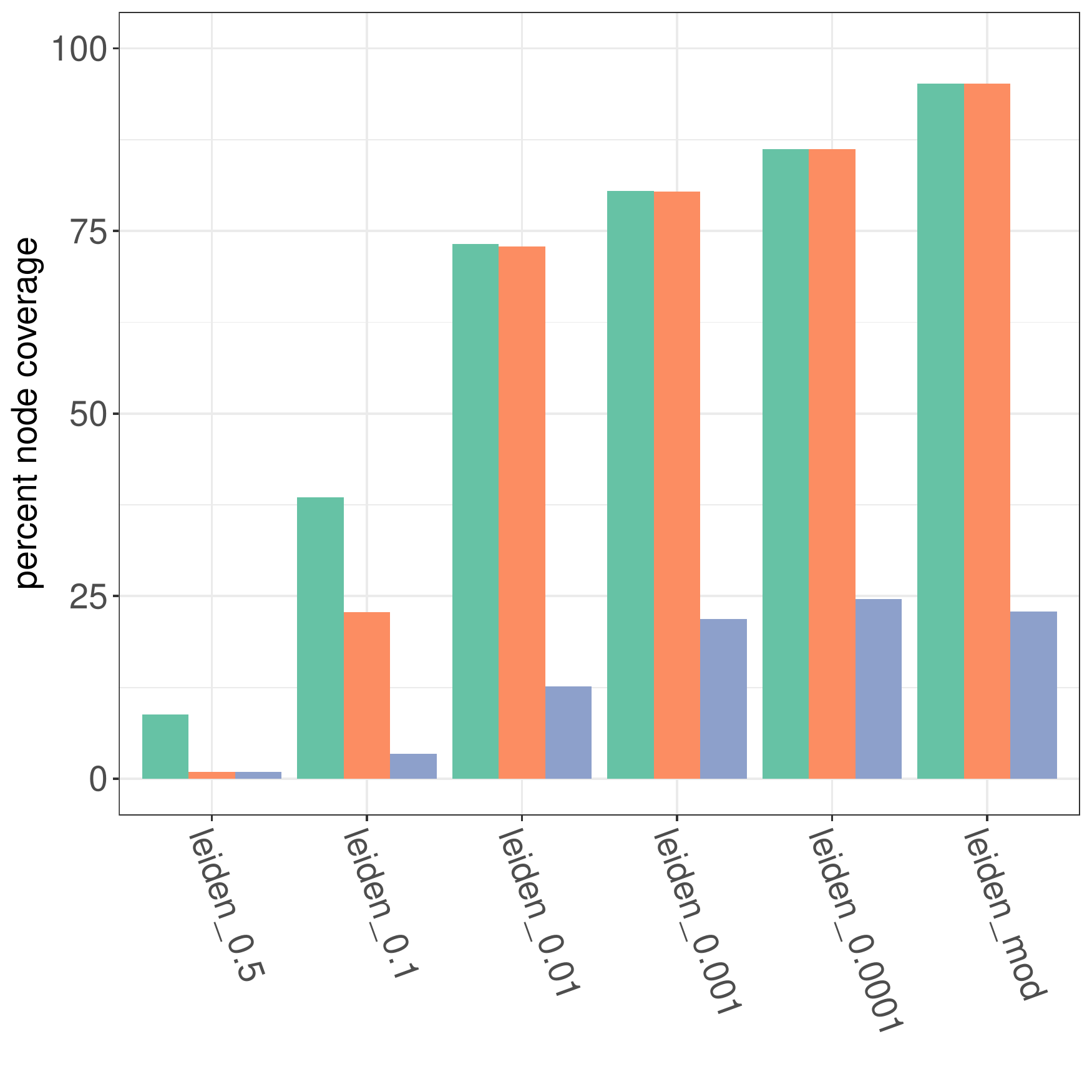}
\caption{CEN}
\end{center}
\end{subfigure}
\caption{\emph {Reduction in node coverage after CM treatment of Leiden clusters.} The Open Citations (left panel) and CEN (right panel) networks were clustered using the Leiden algorithm under CPM at five different resolution values or modularity. Node coverage (defined as
the percentage of nodes in cluster of size at least 2) was computed for Leiden clusters \textcolor{p1}{\large\textbullet} (lime green), Leiden clusters with trees and/or clusters of size 10 or less filtered out  \textcolor{p2}{\large\textbullet} (soft orange),  and after CM treatment of filtered clusters \textcolor{p3}{\large\textbullet} (desaturated blue).}

\label{fig:node-coverage}
\end{figure}

\paragraph{Cluster fate}  To understand the nature of the modifications effected by CM, we further classified the Leiden clusters based on the impact of CM-processing:  \emph{extant, reduced, split, and degraded}, where ``extant'' indicates that the cluster was not modified by CM, ``reduced'' indicates that the cluster is reduced in size, ``split'' indicates that the cluster was divided into at least two smaller clusters, and ``degraded'' indicates that the cluster was reduced to singletons or a cluster of size at most 10. 
We report the fraction in each category (Fig.~5) for six different clustering conditions on the Open Citations and CEN networks. 
When using CM with CPM-optimization with a very large resolution value, most clusters are extant,  the  fraction of extant clusters decreases  for CPM-optimization as the resolution value decreases, and is low also for modularity (Fig.~2).

An interesting trend with split clusters is seen in Figure 5, indicating the initial cluster contained two or more well connected clusters; this is similar to the observation by Fortunato and Barth\'elemy in \cite{fortunato2007resolution} of modularity's behaviour on a ring-of-cliques, where modularity returns clusters that contain two or more cliques, instead of returning the individual cliques,  under some conditions. 
Here we see that clusters that are split by CM unsurprisingly occur in both networks with  modularity, but  also occur in a noticeable way for  CPM-optimization with the smallest resolution value for the Open Citations network. Finally, we note that this occurs for CPM-optimization with other resolution values on both networks (Table S5 and Figure S1).
The most extreme cluster fate is of course when it is degraded to singletons, which occurs in modularity clusterings and CPM-based clusterings at lower resolution values;
however, the degree to which this occurs depends on the network and resolution value in a way that does not suggest any particular pattern (Figure S1).

\begin{figure}[ht!]
\centering
\includegraphics[width=0.85\linewidth]{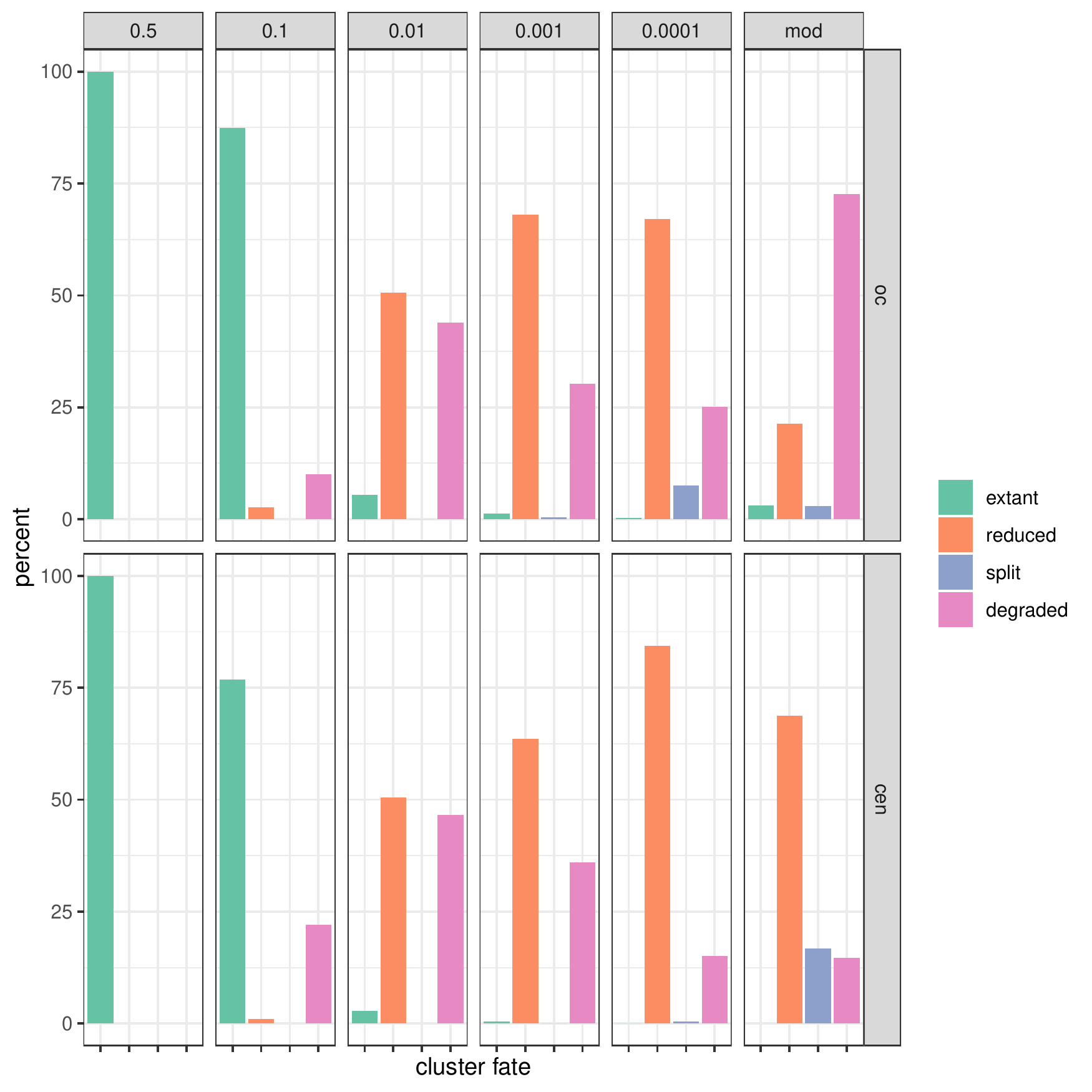}
\caption{\emph{Cluster Fate for Leiden.} CM-processed clusters are classified as extant (unaffected by CM), reduced (replaced by a single smaller cluster), split (replaced by two or more clusters), or degraded (replaced entirely by singletons).  
Top: Cluster Fate for Open Citations (OC); Bottom: Cluster fate for the CEN.
Data are shown for CM treatment of Leiden clusters, under either CPM  for 5 different resolution values or modularity.  
Cluster counts are expressed as a percentage of the input clusters.  }
\label{fig:cluster_fate} 
\end{figure}

\paragraph{Clustering with IKC} 
We examine the impact of CM on IKC (with $k=10$) on the Open Citations and CEN networks (Fig.~6). 
In comparison to Leiden, IKC-clustering results in relatively low node coverage, 23.6\% and 3.8\% in the case of the Open Citations and CEN networks respectively. CM treatment of these clusterings also has a small effect on node coverage.
We also see that most clusters are  extant and some are split, but none are reduced or degraded.

In summary, the response to CM-processing differed across methods. On the CEN and Open Citations networks, and for both IKC and Leiden under modularity or CPM,  node coverage in a post-CM clustering did not exceed 68\%, either because CM-processing reduces node coverage substantially from the initial clustering  in the case of Leiden with modularity or CPM-clustering with smaller resolution values or because the initial clustering was conservative and already had low node coverage. This suggests the possibility that in these real-world networks, only a fraction of the nodes may be in clusters that are sufficiently well connected and sufficiently large.  In other words, real-world networks may not be fully covered by what we consider ``valid'' communities.

\begin{figure}[ht]
\centering
\begin{subfigure}[t]{0.45\textwidth}
\begin{center}
\includegraphics[width=0.85\linewidth]{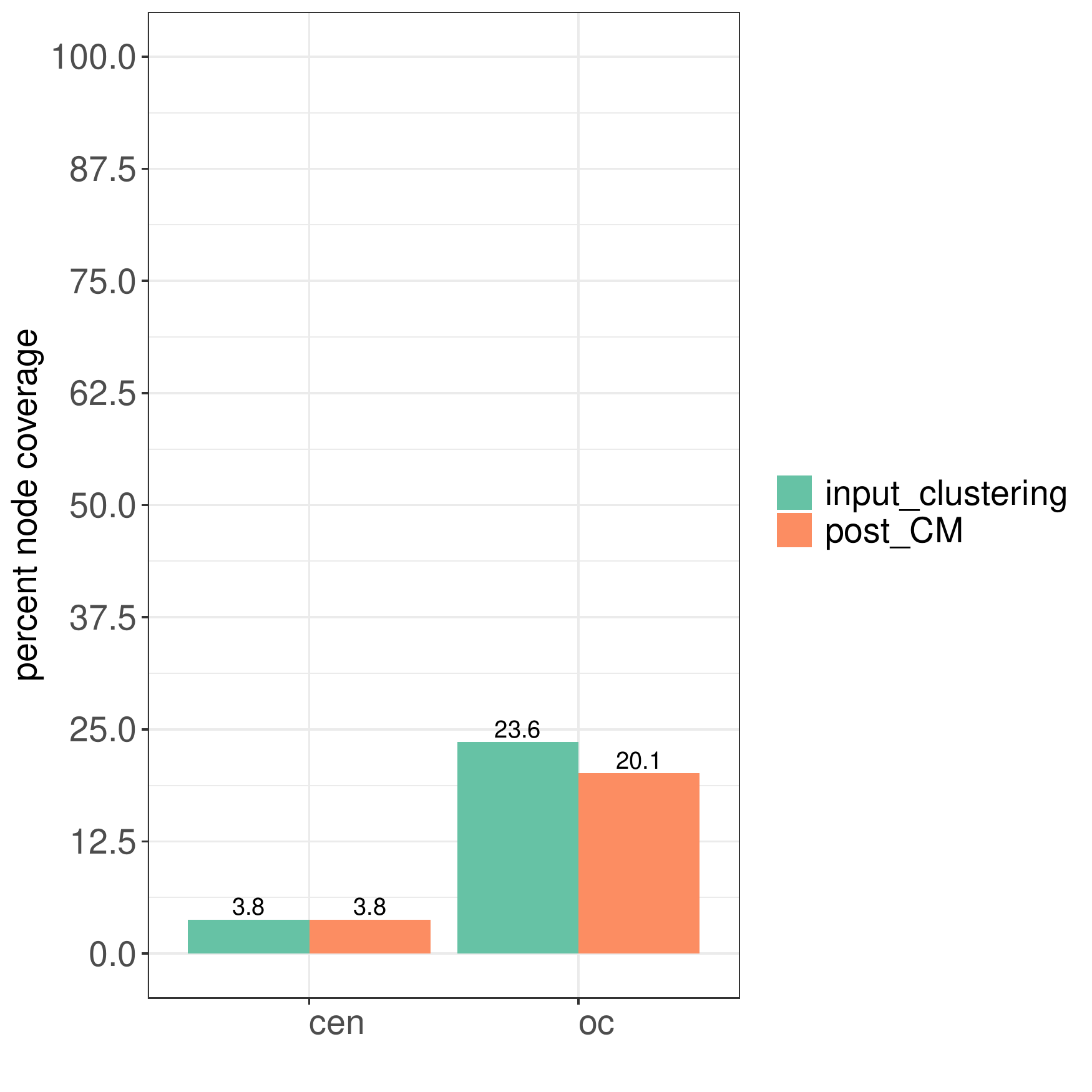}
\caption{Node coverage}
\end{center}
\end{subfigure}
\begin{subfigure}[t]{0.45\textwidth}
\begin{center}
\includegraphics[width=0.85\linewidth]{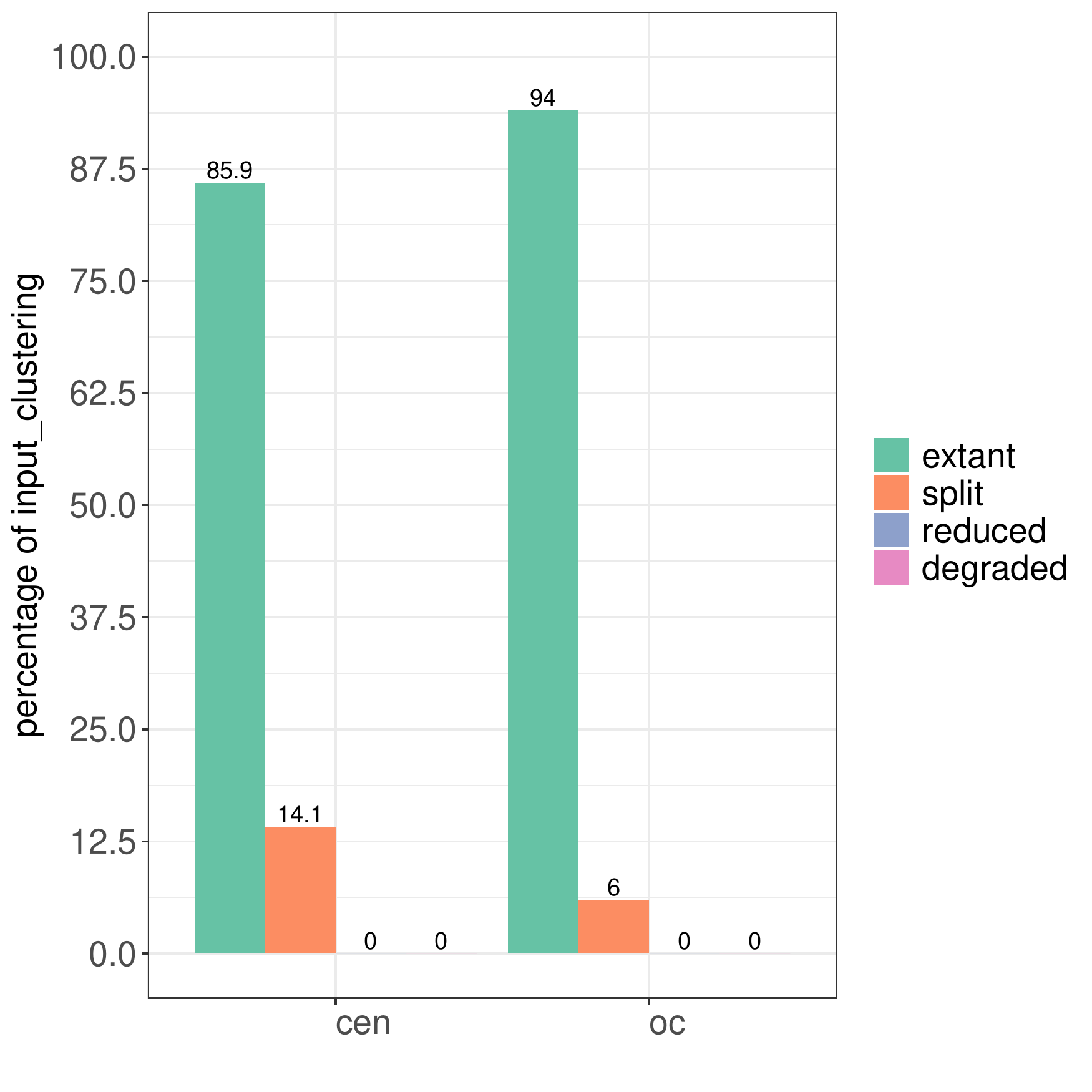}
\caption{Cluster Fate}
\end{center}
\end{subfigure}
\caption{\emph{Node coverage (a) and cluster fate (b) for IKC clusters modified by CM on the Open Citations and CEN.} 
For IKC on these two networks,  node coverage is small on both the CEN and Open Citations, and slightly reduced by CM on the Open Citations network. 
We also see that most clusters are  extant (unaffected by CM-processing) and some are split (replaced by two or more smaller clusters), but none are reduced (replaced by a single smaller cluster) or degraded (replaced by only singletons).
}
\label{fig:is_touched}
\end{figure}

\paragraph{Analysis of synthetic networks} 

\begin{figure}[ht!]
\centering
\includegraphics[width=0.8\linewidth]{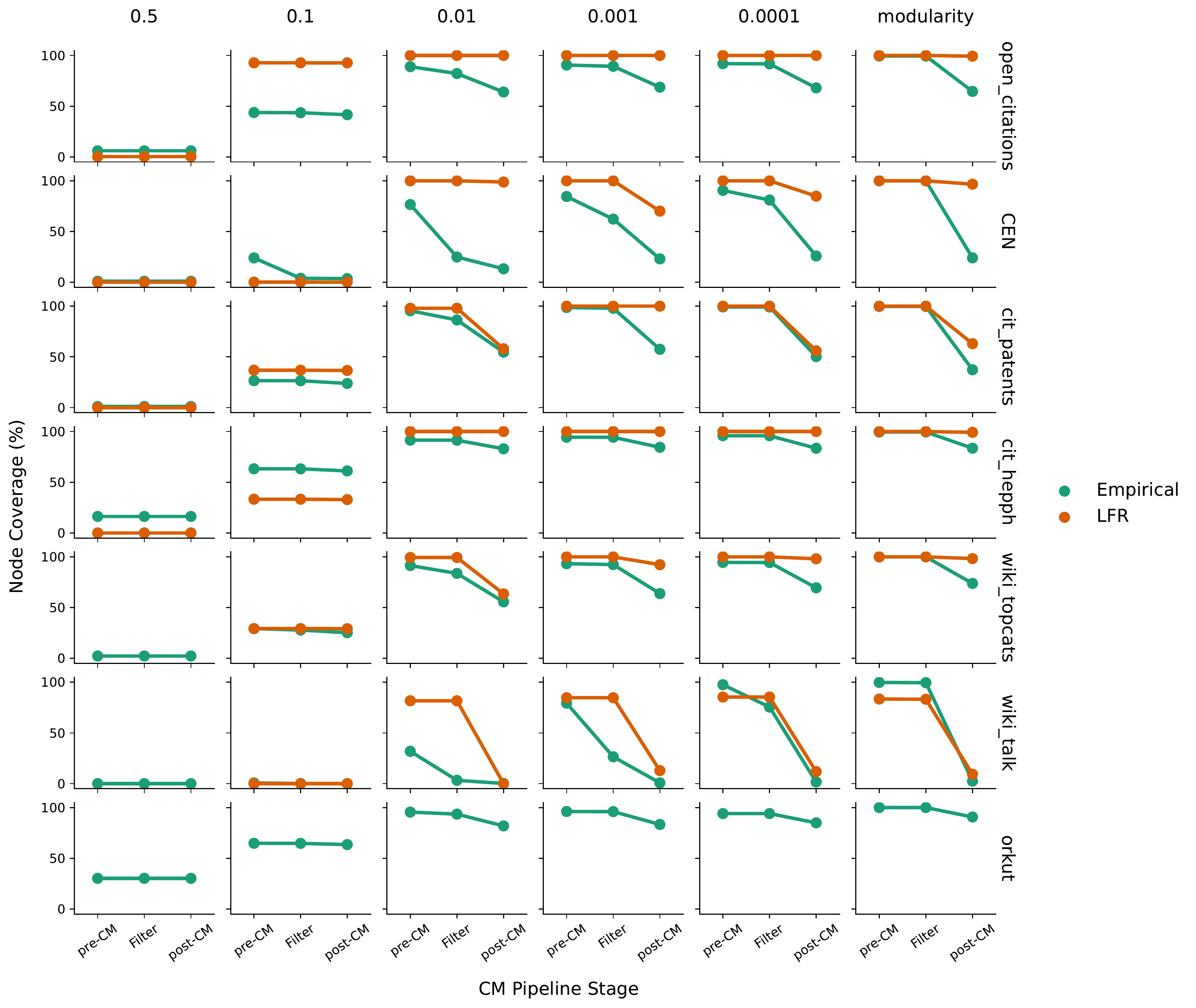}
\caption{\emph{The effect on node-coverage (percentage of nodes in clusters of size at least 11) produced by the CM processing pipeline on LFR and real-world networks.}
 Each row corresponds to a  real-world network, and each column corresponds to a Leiden clustering (either modularity or CPM-optimization with the specified resolution value).
Within each entry, we show node coverage for the Leiden clustering as it goes through the CM pipeline; results on the empirical network are shown in green and results for the LFR network are shown in red.  ``Pre-CM'' indicates the output of the Leiden clustering, ``filter'' is after removing trees and clusters below size 11,  and ``post-CM'' is after the entire pipeline (which also removes any final clusters below size 11). No results are shown for some combinations  (specifically,  data are not shown for LFR for any clustering of the Orkut network nor for the wiki\_topcats and wiki\_talk networks with resolution value $r= 0.5$)  due to LFR failing to generate networks for those settings. }
\end{figure}

\begin{figure}[ht]
\includegraphics[width=0.495\linewidth]{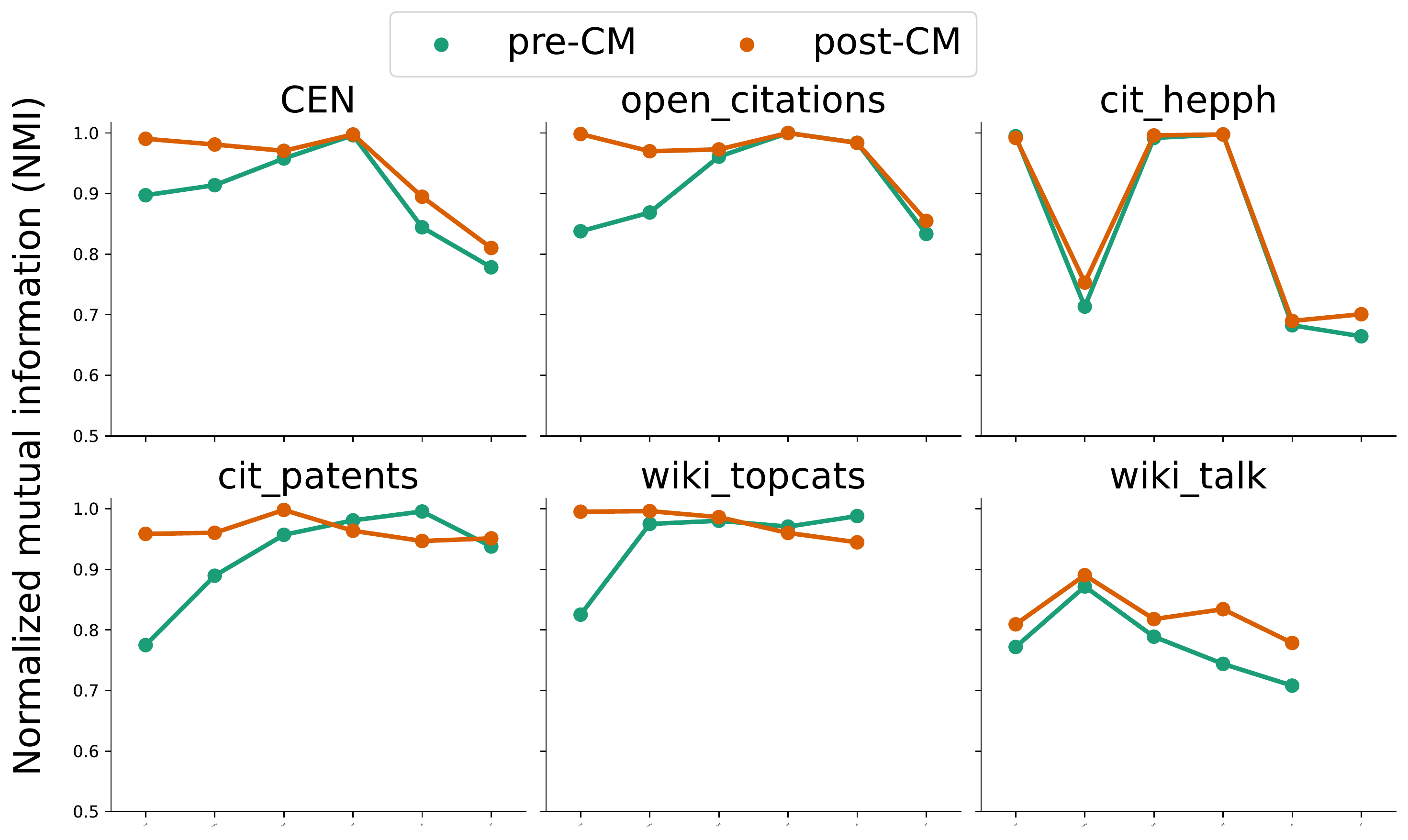}
\includegraphics[width=0.495\linewidth]{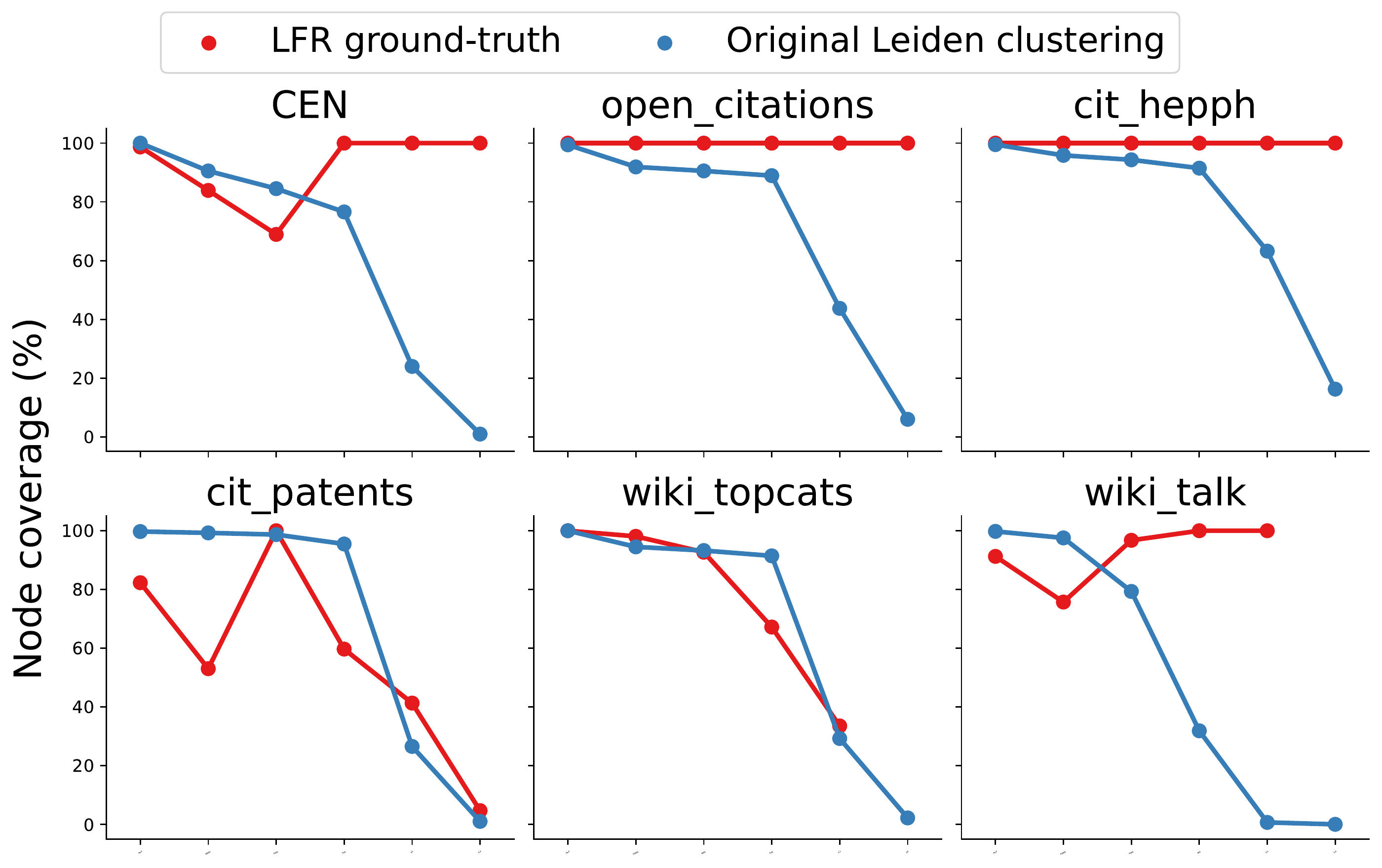}
\includegraphics[width=0.495\linewidth]{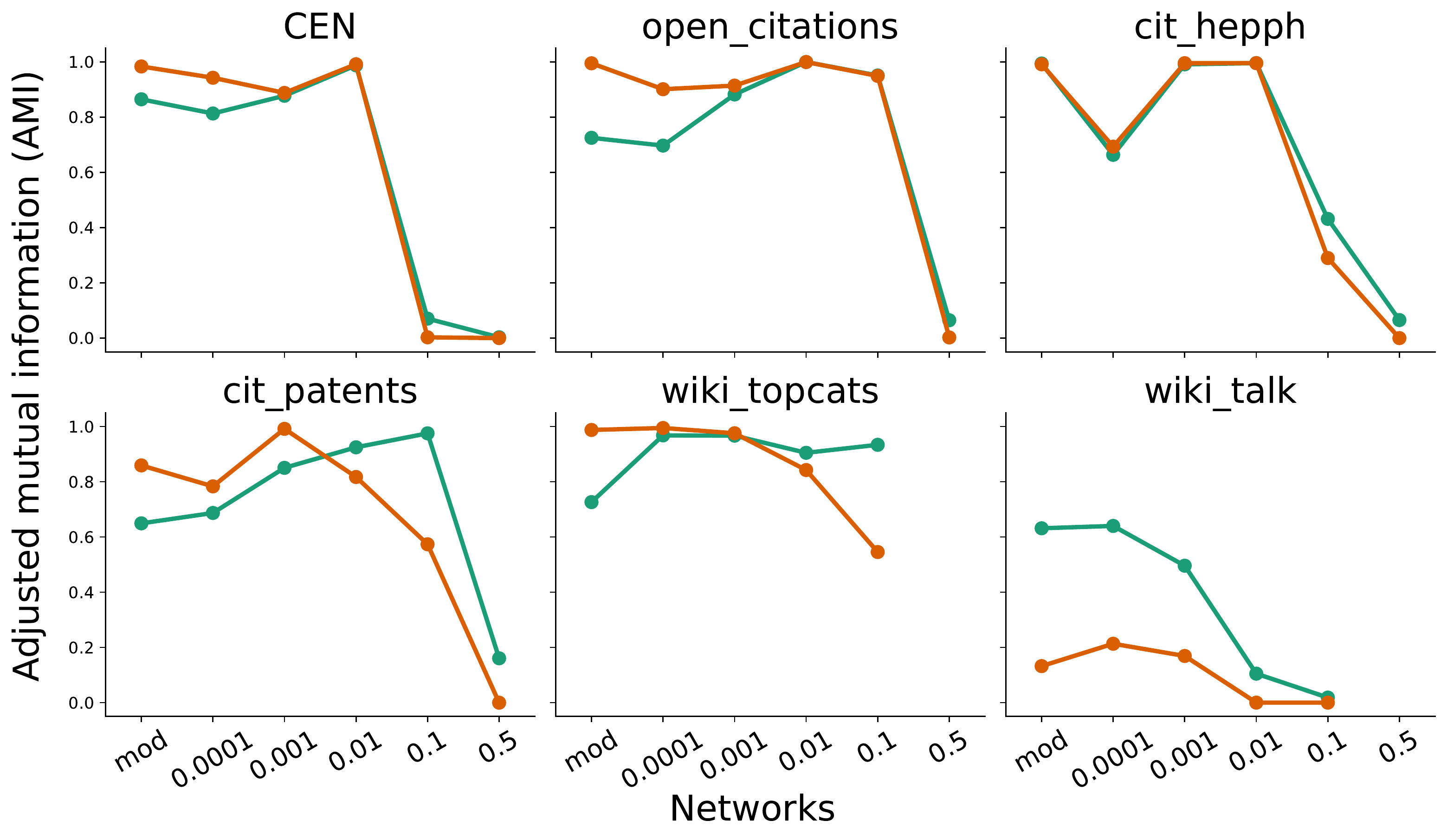}
\includegraphics[width=0.495\linewidth]{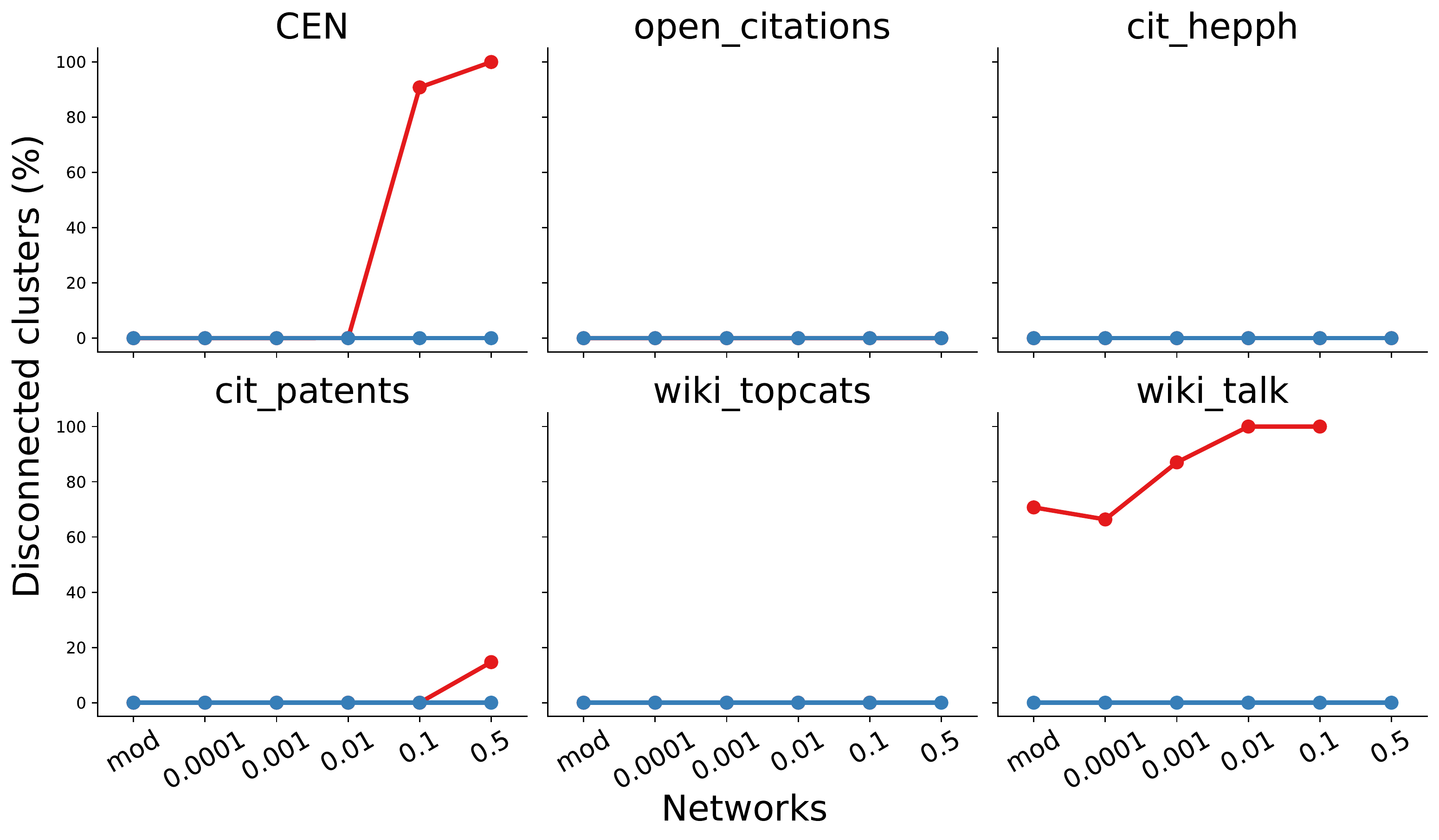}
\caption{\emph{Impact of CM-processing on accuracy of synthetic networks.} The left panels show accuracy measured in terms of NMI and AMI, with respect to the LFR ground-truth communities. Each condition on the x-axis corresponds to a \emph{different} LFR network, generated based on Leiden-modularity or Leiden-CPM with that specific resolution parameter. In total, there are 34 different LFR networks. The right panels show percentage of the LFR ground-truth communities that are small or disconnected. In most conditions, CM improves the accuracy of the original Leiden clustering, except for some of the conditions when the ground-truth communities have many (at least $60\%$) disconnected clusters, or the node coverage by clusters of size at least 11 is low (at most 70\%).}
\label{fig:lfr-accuracy-all-measures}
\end{figure}

Given our hypothesis that the large reduction in node coverage that we observe on real-world networks may be the result of them not being universally covered  by well connected true communities, we examined synthetic networks, noting that 
 synthetic networks generated using LFR software \cite{lancichinetti2008benchmark} assign every node to a non-singleton ground-truth community. 
 If node coverage after CM-processing of LFR networks were similar to real-world networks, it would
 argue against this hypothesis. Examining synthetic networks also enables us to evaluate the impact of CM-processing on accuracy.

For this experiment, we computed  statistics for the Leiden clusterings of the 7 real-world networks we explored, Open Citations, CEN, and 5 networks from the SNAP collection \cite{snapnets}), and used them as input to the LFR software  \cite{lancichinetti2008benchmark} (see Materials and Methods). Using this software, we were not able to build synthetic networks for the CEN and Open Citations with the number of nodes in the empirical networks;  therefore, for these two specific cases, we constructed the LFR networks  to $3$M nodes. For some combinations of network and Leiden clustering, we were unable to produce any synthetic networks, for example, we were unable to produce LFR networks for the Orkut social network. We produced a collection of 34 LFR networks with ground truth communities that had similar empirical statistics as their corresponding real-world networks (Supplementary Materials Section S3). We clustered each of these 34 LFR networks using the same clustering method used to provide empirical statistics to LFR.

Results in Figure 7  show node coverage after clustering for both the empirical network and its corresponding LFR network. The general trends we observed in previous experiments for OC and CEN also hold for the five other empirical networks examined here both with respect to node coverage in Leiden clustering and how CM-processing impacts node coverage.  An interesting trend we see here is that node coverage drops for some CPM clusterings of empirical networks though not for LFR networks after the filtering stage.
Since node coverage in this figure is with respect to clusters of size 11 or larger, any drop in node coverage resulting from filtering is due only to removing trees. 
The large drop in node coverage for Leiden with CPM-optimization in these cases means that CPM-optimization  in some conditions produces a large number of  tree clusters. We note this did not occur for resolution value $r=0.5$, where clusters tend to be small but did occur for other resolution values (Supplementary Tables S3 and S4).

Of specific interest is whether clusterings of the LFR networks respond similarly to CM-processing as clusterings of their corresponding empirical networks, as this addresses our hypothesis regarding why CM-processing impacts node coverage in empirical networks.
While node coverage does drop to the same degree for some LFR networks as for the empirical networks on which they are  based, there  are many cases where node coverage drops much more  for the real-world network than for the  corresponding LFR network, and no cases where there is a bigger drop in node coverage for the LFR network than for the real-world network. Thus, the idea that real-world networks not having universal coverage by valid communities cannot be ruled out.

 We  examined the impact of CM-processing on clustering accuracy; results for Normalized Mutual Information (NMI) and Adjusted Mutual Information (AMI) are shown in Figure 8, and results for the Adjusted Rand Index (ARI) are similar to AMI and are provided in Supplementary Materials, Figure S11 \cite{danon2005comparing,vinh2009information,hubert1985comparing}. We see that CM-processing improves NMI accuracy for modularity and also for CPM-optimization when used with small resolution values. CM-processing tends to be neutral for NMI otherwise, and was only detrimental on  two network:clustering pairs. The impact on AMI accuracy is more variable. For example, CM-processing reduced AMI accuracy for all wiki\_talk network:clustering pairs except for CPM-optimization with
 $r=0.1$ where accuracy was very low and the impact was neutral. CM-processing also reduced accuracy for CPM-optimization on some network:clustering pairs for large resolution values.
 
However, when examining the properties of the ground-truth clusterings of these networks, we see  that the cases where CM-processing produced a noteworthy reduction in  accuracy for NMI or AMI are those where there are many disconnected ground truth clusters, for example all  wiki\_talk clusters, or where the node coverage by clusters of size at least 11 is small, for example cit\_patents at the three largest resolution values and wiki\_topcats with the two largest resolution values. However, there are conditions with low node coverage or a large fraction of disconnected clusters where CM-processing is neutral or even beneficial.

The occurrence of disconnected ground-truth clusters in the LFR networks is striking and problematic, since a basic expectation of a community is that it is connected, if not well-connected \cite{traag2019louvain}.
Hence, we assert that it is unreasonable to evaluate accuracy with respect to a ground-truth set of communities if the communities are not connected.
In other words,  it does not make sense to evaluate clustering accuracy for those LFR networks that contain many disconnected ground truth communities, including
the entire set of LFR networks constructed for wiki\_talk invalid, which is one of the conditions where CM-processing reduces AMI accuracy. The fact that LFR networks had ground truth clusters that were not connected also indicates the failure of LFR software to reproduce features of the input network:clustering pairs, which by construction always have 100\% of the clusters connected.

It is easy to see why a low node coverage by clusters of size at least 11 could reduce accuracy for CM-processing, since CM automatically removes all small clusters.
However, this property depends on the network:clustering pair, which depends on network features as well as the clustering methodology.  In this study, clustering real-world networks produced a large fraction of small clusters when we used CPM-optimization with large resolution values; the conditions where CM-processing reduced AMI accuracy on the LFR networks with low node coverage by clusters of size at least 11 are drawn from those conditions.
We also  note that users typically select the clustering method that produces cluster sizes that match their interest. Therefore, CM-processing will not be beneficial where there is interest in recovering small communities unless the bound $B$ is replaced by a smaller value.

In interpreting these results, we also note the discrepancy between some empirical statistics of LFR networks and those of the real-world network:clustering pairs that were used to simulate the LFR network.  Beyond incomplete matching of features,
differences such as the frequency of disconnected clusters and the percentage of clusters that are small make it questionable whether accuracy on LFR networks is suggestive of accuracy on real-world networks.

\paragraph{Discussion} In this study, we considered the question of whether clusters produced by community detection methods are well connected. We use a mild definition to demonstrate that five different clustering paradigms generate, to varying extents, output clusters that are not well connected. An important implication of these results is that portions of a network may not exhibit community structure. Further, at parameter settings that maximize node coverage, weakly connected parts of a network may be forced into communities. Related prior work \cite{Miasnikof2023,gao1710.00862} address  whether a graph in general is clusterable, which is a related question.

We developed CM to convert, through partitioning,  poorly connected clusters into well connected ones. For flexibility, the function in CM that defines ``well connected'' can be modified by users to be more or less stringent.  Similarly, we provide a parameter $B$, tunable by the user, that specifies the smallest size of a cluster for it to be retained; this too can be modified by the user to be more or less stringent.  At present, CM provides support only for Leiden and IKC, the most scalable of the methods we tested. CM is being extended to provide support for Infomap and MCL and concurrently being redesigned for developers to integrate their own clustering methods into it. CM allows the user to explore cluster quality in the input, as it reveals which clusters are poorly connected, and, in some cases, finds substructure within clusters. Thus, CM-processing can be used to evaluate and improve clustering outputs, and interrogate and explore the community structure within a given network.

Several factors affect how significantly  CM-processing changes a given clustering.  These include the network itself, as some networks seem to be more impacted by CM-processing. We also see that the choice of  resolution parameter for Leiden-CPM has an influence on how much the clustering changes, with generally larger impact for small resolution values. 

The findings that LFR networks produce different patterns than empirical networks is perhaps not surprising, but here we consider potential explanations. First, the LFR methodology assumes that the degree distribution and cluster size distributions follow a power law, however, it is not clear that this is universal to citation and real-world networks \cite{Radicchi2008,Stringer2010,artico2020rare,brzezinski2015power}. 
Furthermore, we also observed that the degree distribution and cluster size distributions were imperfectly fit by LFR software in our study (Supplementary Materials, Figs. S2 and S3). 
Hence, the assumptions about node degree and cluster size distributions that govern the LFR model may not result in adequate simulation of real-world networks. 

Another assumption in the LFR methodology is that every node is in a community, which is one that would benefit from deeper investigation.
Intuitively, we posit that the assumption that every node in a real world community is in a community may only be reasonable if the communities can be small and/or  poorly connected. 

We recognize that our perspective on well connected clusters may result in narrow descriptions of communities, perhaps akin to cores of core-periphery structures \cite{Breiger2014,Rombach2017,Wedell2022}. 
Additionally, informative weaker links \cite{granovetter1973strength} may be lost from communities since CM partitions input clusters that are poorly connected into sets of well connected clusters.  However, such weak links are not lost from the network being analyzed. 

Having developed CM and ascertained its quantitative effects, a direction for our future 
work is to evaluate it in the context of specific evaluations that are supported by mixed methods. More generally, we emphasize leaving the definition, use, and interpretation of well connected to users. 

\vspace{.1in}

\section*{Materials and Methods}

\paragraph{Defining well connected clusters.} 

\begin{figure}[ht]
\centering
\begin{center}
\includegraphics[width=0.5\textwidth]{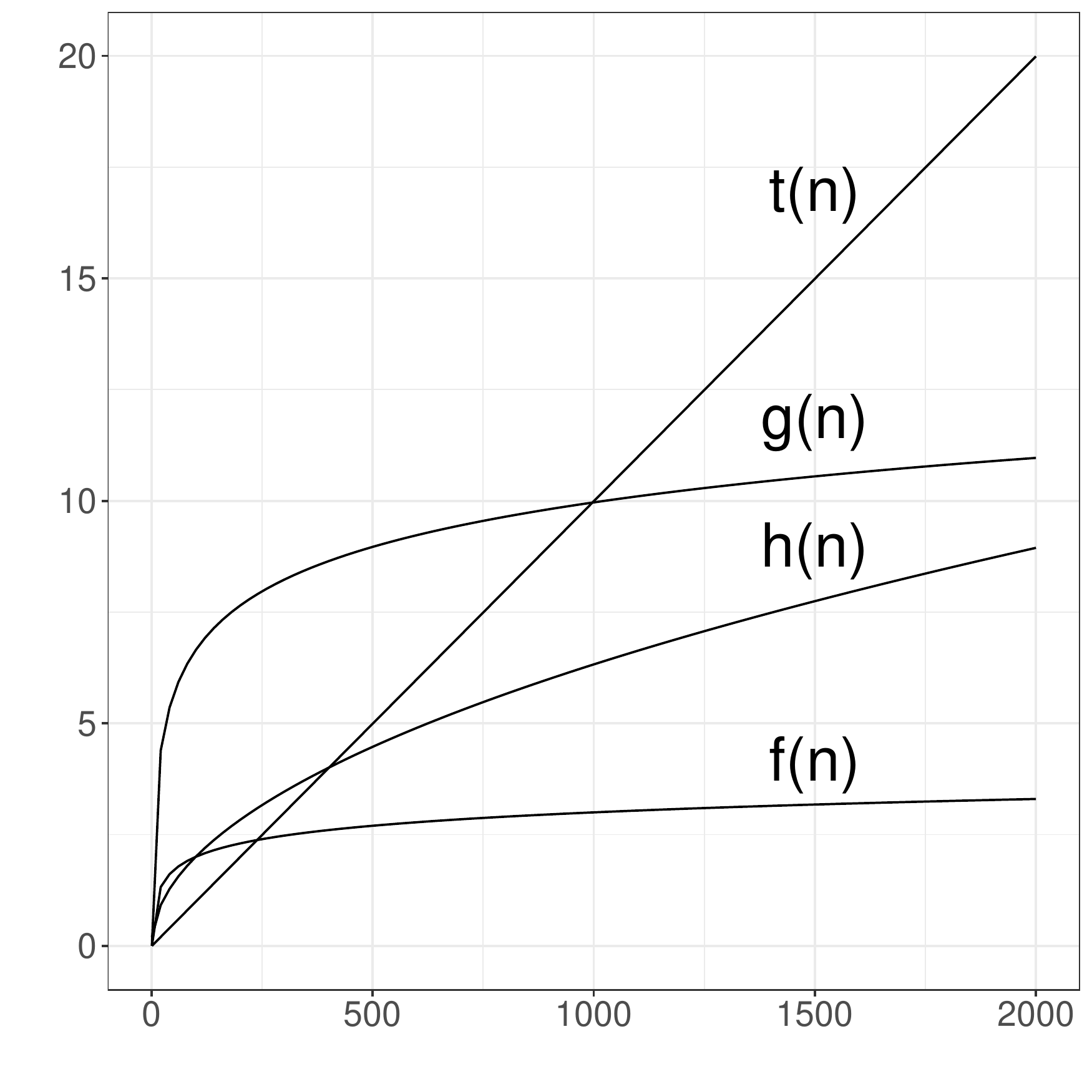}
\end{center}
\caption{\emph{Comparison of the lower bounds on edge cut sizes  for defining well connected clusters. } Here we compare four functions; 
$t(n) = 0.01 (n-1)$ is Traag's function, which provides a  lower bound of the cut size for a cluster on $n$ nodes in  a CPM-optimal clustering when $r=0.01$, $f(n) = \log_{10}n$, $g(n) = \log_2 n$, and $h(n) = \frac{\sqrt n}{5}$. For $n \geq 1000$, Traag's function provides the largest lower bound on the edge cut size, and
so is the strongest guarantee of the functions we compare, while $f(n)$ provides the smallest lower bound for all $n \geq 239$. 
However, for $n \leq 238$, $f(n) \geq t(n)$, so that $f(n)$ provides a stronger guarantee on these small- to moderate-sized clusters than Traag's bound.
 }
\label{fig:prelim-study-lowerbounds}
\end{figure}

In \cite{traag2019louvain}, Traag et al.~proved that every cluster $C$ in a CPM-optimal clustering of a network (with resolution parameter $r$) would satisfy the following property: any edge cut $X$ of $C$ 
splitting the nodes into two sets $A$ and $B$ would have at least $r \times |A| \times |B|$ edges.
This function is maximized when $|A|=|B|$ and minimized when $|A|=1$ and $|B|=n-1$, where $C$ has $n$ nodes.
Hence in particular, this establishes that the minimum cut  for any cluster with $n$ nodes has size at least $r \times (n-1)$.

We have already observed that this bound  (which we refer to as Traag's bound) is weak when $n$ is not large and $r$ is very small. 
For example, when $r=0.01$ and $n=50$, the bound only establishes that the minimum edge cut is at least $1$, which therefore simply asserts that the
cluster is connected. 
Hence, we seek lower bounds on the size of the minimum edge cut that are larger than Traag's bound of  $r \times (n-1)$ for small values of $n$ but grow very slowly, and so do not exceed 
Traag's bound for larger $n$.

In the text, we discussed the bound that requires that for a cluster of $n$ nodes to be well connected,   the size of its min cut must be {\em strictly greater} than $f(n) = \log_{10} n$. 
We compare $f(n)$ to Traag's lower bound when $r=0.01$, which we refer to as $t(n)$, as well as two other functions, $g(n) = \log_2 n$ and $h(n) = \frac{\sqrt n}{5}$;
note that each of $f(n), g(n)$ and $h(n)$    is strictly positive and increasing, and yet grows more slowly than  Traag's function $t(n)$.

The comparison between these functions in the range of cluster sizes $1 \leq n \leq 2000$ is shown in Figure 9.
As desired, for large enough $n$, Traag's function $t(n)$ dominates the other functions, thus imposing a much stronger guarantee on the
minimum cut size for the cluster.
We also note that $f(n) < g(n)$ for all $n>1$, but that the relationship between $f(n), h(n)$, and $t(n)$  depends on $n$.
For large $n$, however, $f(n)$ is less than all the other functions, making it the most slowly growing function.

Based on this comparison, we used $f(n)$ to define when   a cluster of $n$ nodes is well connected: the min cut size is strictly greater than $\log_{10}(n)$; otherwise
we say that the cluster is poorly connected. 

Note that if we had picked  $g(n)$ instead,  we would have considered more clusters poorly connected, since $f(n) < g(n)$ for all $n$.
Similarly, picking $h(n)$ would have generally been more stringent a requirement (at least for all but very small values of $n$), and so would have
ruled more clusters as being poorly connected.
Thus, $f(n)$ represents a very modest constraint on the size of the min cut for a cluster, in order for it to qualify as being well connected.

\paragraph{The Connectivity Modifier (CM) Pipeline}  To remediate poorly-connected clusters, we developed a modular pipeline that we now describe (Fig. 2). By design, this pipeline is guaranteed to return a clustering where each cluster is well connected according to the function $f(n)$  and has size at least $B$.  The values of $f(n)$ and B used in this study are set as default but can be easily modified by a user. Note that if an input cluster meets these two criteria, it will be present in the output clustering.  Furthermore, every cluster in the output will either be one of the input clusters or will be a subset of one of the input clusters. 

The CM pipeline requires the user to specify values for three algorithmic parameters:
\begin{itemize}
\item $B$, the minimum allowed size of any ``valid" vertex community (default $B=11$)
\item $f(n)$, so that a cluster is well connected only if its min cut size exceeds $f(n)$ (default: $f(n)= \log_{10} n$)
\item A clustering method, presently selected from Leiden optimizing CPM, Leiden optimizing modularity, or the Iterative k-core (IKC) method with $k=10$. Support for additional methods is in active development.
\end{itemize}

\noindent
The input to the CM pipeline is a network $\mathcal{G}$ with $N$ nodes and the algorithmic parameters as specified.
The pipeline then operates in four stages:
\begin{itemize}
\item
Stage 1: A clustering is generated  from the input network $\mathcal{G}$.
\item Stage 2: The clustering is filtered  to remove clusters below size $B$ and all trees, which are not well-connected when using $f(n) = \log_{10} n$
\item Stage 3: The Connectivity Modifier (CM) is applied to each cluster that remains.  
First, all nodes of degree at most $f(n)$ are removed (where $n$ is the number of nodes in the cluster), until there are no low degree nodes remaining.
Then, 
for each remaining cluster, a min cut is calculated using VieCut \cite{henzinger2018practical}, and if it is not greater than $f(n)$  in size then the min cut  is removed, thus splitting the cluster into two components.    
These components are then re-clustered using the selected clustering method, and the process repeats until all clusters are well-connected.
\item Stage 4: Any resultant clusters below size $B$ are removed.
\end{itemize}

\paragraph{Additional details for the Connectivity Modifier} The Connectivity Modifier was developed using Python 3.10. The original implementation of VieCut \cite{henzinger2018practical} in C++17 was wrapped into a Python package using PyBind11. In our analysis we use unbalanced minimum cuts computed with the NOI algorithm \cite{nagamochi94}.

\paragraph{Additional details for IKC}
Due to memory constraints, IKC fails to run with the connectivity modifier on the entirety of the OpenCitations network. To solve this, we modified IKC to run in memory as an imported Python module rather than as a separate executable. 
Moreover, we split the initial IKC clustering of OpenCitations into separate clusters, and the Connectivity Modifier was then run on the largest clusters independently (see Supplementary Section S2 for full details).
\paragraph{Data } 

A custom-implemented Extract-Transform-Load (ETL) process was designed to process the publicly available Open Citations dataset, downloaded in Aug 2022,  and load it into a PostgreSQL table. The resultant network contained 75,025,194 nodes and 1,363,605,603 edges. The CEN is a citation network constructed from the literature on exosome research. From the SNAP repository, we downloaded cit\_hepph,  an  Arxiv High Energy Physics paper citation network; cit\_patents, a citation network among US patents; orkut, a social media network; wiki\_talk, a network containing users and discussion from the inception of Wikipedia until January 2008; and wiki\_topcats, a web graph of Wikimedia hyperlinks. Before clustering, all networks were processed to remove self-loops as well as duplicate and parallel edges.

\begin{table}[ht]
\centering
\begin{tabular}{lrrrr}
  \hline
 network & nodes & edges & avg\_deg & ref \\
  \hline
Open Citations &   75,025,194 & 1,363,605,603  &  36.35  &   \cite{Peroni2020}\\
CEN			&  13,989,436 &  92,051,051 & 13.16 & \cite{Jakatdar2022}\\
cit\_hepph 	&   34,546 & 420,877 & 24.37 &  \cite{Leskovec2005}\\ 
cit\_patents 	&   3,774,768 & 16,518,947 &   8.75  &  \cite{Leskovec2005}\\
orkut 		& 3,072,441 & 117,185,083 & 76.28 &  \cite{Yang2013}\\
wiki\_talk 		&   2,394,385 & 4,659,565 & 3.89 & \cite{Leskovec2010}  \\
wiki\_topcats 	&   1,791,489 &  25,444,207 & 28.41 & \cite{Yin2017}\\
   \hline
\end{tabular}
\caption{Summary statistics for networks used in this study. Average degree is the average of the node degrees across the network.}
\label{tab:empirical-stats-all}
\end{table}

\paragraph{LFR (synthetic) networks}

To create simulated networks with ground truth communities, while attempting to emulate the properties of each empirical network and its corresponding Leiden clustering, we used the LFR software from \cite{lancichinetti2008benchmark}. The generative model of the LFR graphs assume that the node degree and the community size distributions are power-law distributions \cite{albert2002statistical}.
The software for generating LFR benchmark graphs \cite{fortunato-resources} takes the following eight parameters as input:
\begin{itemize}
    \item  Network properties: Number of nodes $N$, average and maximum node degrees ($k$ and $k_{max}$ respectively), and negative exponent for degree sequence ($\tau_1$) that is assumed to be a power-law.
    \item Community properties: Maximum and minimum community sizes ($c_{max}$ and $c_{min}$), and negative exponent for the community size distribution ($\tau_2$), also modeled as a power-law.
    \item Mixing parameter $\mu$, that is the ratio between the degree of a node outside its community and its total degree, averaged over all nodes in the network. Lower $\mu$ values suggest that a network consists of well-separated communities, as nodes are mostly connected to other nodes inside their communities, rather than outside of it.
\end{itemize}

\paragraph{Parameter Estimation.} To emulate the empirical networks using LFR graphs, we estimated all eight parameters described above for a given pair of network $\mathcal{G}$ and a clustering $\mathcal{C}$. Computing $N, k, k_{max}, c_{min}$ and $c_{max}$ is straightforward using \emph{networkX} \cite{hagberg2008exploring,lfr-generation-code}. To estimate $\mu$, we perform a single iteration over all edges of the network, and for each edge, if the nodes on the two sides of it were in different communities, that edge contributes to the ratio $\mu$ of these two nodes. The total $\mu$ of the network:clustering pair is the average $\mu$ across all the nodes.

To estimate $\tau_1$ and $\tau_2$, we fit a power-law distribution to the node degree sequence and the community size distribution, using the approach from \cite{clauset2009power} that is implemented in the \emph{power-law} Python package \cite{alstott2014powerlaw}. Note that the power-law property may hold for the tail of the degree or community size sequence and not the whole distributions. Therefore, following \cite{clauset2009power}, we estimate $x_{min}$, the minimum value for which the power-law property holds as well as the exponent $\alpha$ for the tail of the distribution.

\paragraph{Generating LFR networks.}

After computing these parameters based on the Leiden clusterings of the empirical networks using both modulatiry and CPM with a range of  resolution parameters, we simulated LFR networks using the software from \cite{lancichinetti2008benchmark}, producing empirical statistics reported in Supplementary Tables S7 and S8.

For networks with more than 10 million nodes, i.e., Open Citations and the CEN, we limited the number of vertices to 3 million, due to scalability limitations of  the LFR benchmark graph generator \cite{slota2019scalable}, while preserving the edge density reflected by average degree, and the mixing parameter.  
The numbers of nodes of the  other LFR graphs exactly match the number of nodes in the corresponding empirical network. In some cases, due to the inherent limitations of the LFR graph generator, we had to modify the ranges of the community sizes, i.e., increase $c_{min}$ and decrease $c_{max}$, to generate the network.
These values are available with the datasets.

As shown in the statistics reported in Supplementary Tables S7 and S8, the average node degree, the mixing parameter, and the exponents for the degree and community size distributions are in all cases very well-preserved. Further details about the pipeline for producing LFR graphs and the statistics of the graphs are provided in Supplementary Materials Section 3. We calculated NMI, AMI, and ARI using the Python Scikit-Learn package \cite{pedregosa2011scikit}. Other software used includes Leiden and IKC \cite{leiden-code,ikc-code}.

\bibliography{cm_sa_rev3_arxiv.bib}

\begin{thebibliography}{10}

\bibitem{newman2004finding}
M.~E. Newman, M.~Girvan, Finding and evaluating community structure in
  networks.
\newblock {\it Physical review E\/} {\bf 69}, 026113 (2004).

\bibitem{Mucha2010}
P.~J. Mucha, T.~Richardson, K.~Macon, M.~A. Porter, J.-P. Onnela, Community
  structure in time-dependent, multiscale, and multiplex networks.
\newblock {\it Science\/} {\bf 328}, 876--878 (2010).

\bibitem{Fortunato2022}
S.~Fortunato, M.~E.~J. Newman, 20 years of network community detection.
\newblock {\it Nature Physics\/} {\bf 18}, 848--850 (2022).

\bibitem{Wedell2022}
E.~Wedell, M.~Park, D.~Korobskiy, T.~Warnow, G.~Chacko, Center-periphery
  structure in research communities.
\newblock {\it Quantitative Science Studies\/} {\bf 3}, 289--314 (2022).

\bibitem{Boyack2013}
K.~W. Boyack, R.~Klavans, Creation of a highly detailed, dynamic, global model
  and map of science.
\newblock {\it Journal of the Association for Information Science and
  Technology\/} {\bf 65}, 670--685 (2013).

\bibitem{Waltman2012}
L.~Waltman, N.~J. van Eck, A new methodology for constructing a
  publication-level classification system of science.
\newblock {\it Journal of the American Society for Information Science and
  Technology\/} {\bf 63}, 2378--2392 (2012).

\bibitem{Boyack2011}
K.~W. Boyack, D.~Newman, R.~J. Duhon, R.~Klavans, M.~Patek, J.~R. Biberstine,
  B.~Schijvenaars, A.~Skupin, N.~Ma, K.~B\"{o}rner, Clustering more than two
  million biomedical publications: Comparing the accuracies of nine text-based
  similarity approaches.
\newblock {\it {PLoS} {ONE}\/} {\bf 6}, e18029 (2011).

\bibitem{Coscia2011}
M.~Coscia, F.~Giannotti, D.~Pedreschi, A classification for community discovery
  methods in complex networks.
\newblock {\it Statistical Analysis and Data Mining\/} {\bf 4}, 512--546
  (2011).

\bibitem{Bonchi2021}
F.~Bonchi, D.~Garc{\'{\i}}a-Soriano, A.~Miyauchi, C.~E. Tsourakakis, Finding
  densest k-connected subgraphs.
\newblock {\it Discrete Applied Mathematics\/} {\bf 305}, 34--47 (2021).

\bibitem{traag2019louvain}
V.~A. Traag, L.~Waltman, N.~J. Van~Eck, {From Louvain to Leiden: guaranteeing
  well-connected communities}.
\newblock {\it Scientific reports\/} {\bf 9}, 1--12 (2019).

\bibitem{Blondel2008}
V.~D. Blondel, J.-L. Guillaume, R.~Lambiotte, E.~Lefebvre, Fast unfolding of
  communities in large networks.
\newblock {\it Journal of Statistical Mechanics: Theory and Experiment\/} {\bf
  2008}, P10008 (2008).

\bibitem{Traag2011}
V.~A. Traag, P.~V. Dooren, Y.~Nesterov, Narrow scope for resolution-limit-free
  community detection.
\newblock {\it Physical Review E\/} {\bf 84} (2011).

\bibitem{Rosvall2008}
M.~Rosvall, C.~T. Bergstrom, Maps of random walks on complex networks reveal
  community structure.
\newblock {\it Proceedings of the National Academy of Sciences\/} {\bf 105},
  1118--1123 (2008).

\bibitem{VanDongen2008}
S.~V. Dongen, Graph clustering via a discrete uncoupling process.
\newblock {\it {SIAM} Journal on Matrix Analysis and Applications\/} {\bf 30},
  121--141 (2008).

\bibitem{Miasnikof2023}
P.~Miasnikof, A.~Y. Shestopaloff, A.~Raigorodskii, Statistical power, accuracy,
  reproducibility and robustness of a graph clusterability test.
\newblock {\it International Journal of Data Science and Analytics\/} {\bf 15},
  379--390 (2023).

\bibitem{Peroni2020}
S.~Peroni, D.~Shotton, {OpenCitations}, an infrastructure organization for open
  scholarship.
\newblock {\it Quantitative Science Studies\/} {\bf 1}, 428--444 (2020).

\bibitem{cm2023}
V.~Ramavarapu, F.~Ayres, M.~Park, V.~K. Pailodi, G.~Chacko, T.~Warnow,
  Connectivity modifier,
  \url{https://github.com/illinois-or-research-analytics/cm_pipeline} (2023).

\bibitem{cm2022}
B.~Liu, M.~Park, Connectivity modifier,
  \url{https://github.com/RuneBlaze/connectivity-modifier} (2022).

\bibitem{fortunato2007resolution}
S.~Fortunato, M.~Barthelemy, Resolution limit in community detection.
\newblock {\it Proceedings of the {N}ational {A}cademy of {S}ciences\/} {\bf
  104}, 36--41 (2007).

\bibitem{lancichinetti2008benchmark}
A.~Lancichinetti, S.~Fortunato, F.~Radicchi, Benchmark graphs for testing
  community detection algorithms.
\newblock {\it Physical review E\/} {\bf 78}, 046110 (2008).

\bibitem{snapnets}
J.~Leskovec, A.~Krevl, {SNAP D}atasets: {Stanford} large network dataset
  collection (2014). Http://snap.stanford.edu/data.

\bibitem{danon2005comparing}
L.~Danon, A.~Diaz-Guilera, J.~Duch, A.~Arenas, Comparing community structure
  identification.
\newblock {\it Journal of statistical mechanics: Theory and experiment\/} {\bf
  2005}, P09008 (2005).

\bibitem{vinh2009information}
N.~X. Vinh, J.~Epps, J.~Bailey, {\it Proceedings of the 26th annual
  international conference on machine learning\/} (2009), pp. 1073--1080.

\bibitem{hubert1985comparing}
L.~Hubert, P.~Arabie, Comparing partitions.
\newblock {\it Journal of classification\/} {\bf 2}, 193--218 (1985).

\bibitem{gao1710.00862}
C.~Gao, J.~Lafferty, Testing for global network structure using small subgraph
  statistics (2017).

\bibitem{Radicchi2008}
F.~Radicchi, S.~Fortunato, C.~Castellano, Universality of citation
  distributions: Toward an objective measure of scientific impact.
\newblock {\it Proceedings of the National Academy of Sciences\/} {\bf 105},
  17268--17272 (2008).

\bibitem{Stringer2010}
M.~J. Stringer, M.~Sales-Pardo, L.~A.~N. Amaral, Statistical validation of a
  global model for the distribution of the ultimate number of citations accrued
  by papers published in a scientific journal.
\newblock {\it Journal of the American Society for Information Science and
  Technology\/} {\bf 61}, 1377--1385 (2010).

\bibitem{artico2020rare}
I.~Artico, I.~Smolyarenko, V.~Vinciotti, E.~C. Wit, How rare are power-law
  networks really?
\newblock {\it Proceedings of the Royal Society A\/} {\bf 476}, 20190742
  (2020).

\bibitem{brzezinski2015power}
M.~Brzezinski, Power laws in citation distributions: evidence from scopus.
\newblock {\it Scientometrics\/} {\bf 103}, 213--228 (2015).

\bibitem{Breiger2014}
R.~Breiger, {\it Explorations in Structural Analysis ({RLE} Social Theory)\/}
  (Routledge, 2014).

\bibitem{Rombach2017}
P.~Rombach, M.~A. Porter, J.~H. Fowler, P.~J. Mucha, Core-periphery structure
  in networks (revisited).
\newblock {\it {SIAM} Review\/} {\bf 59}, 619--646 (2017).

\bibitem{granovetter1973strength}
M.~S. Granovetter, The strength of weak ties.
\newblock {\it American {J}ournal of {S}ociology\/} {\bf 78}, 1360--1380
  (1973).

\bibitem{henzinger2018practical}
M.~Henzinger, A.~Noe, C.~Schulz, D.~Strash, Practical minimum cut algorithms.
\newblock {\it {ACM} Journal of Experimental Algorithmics\/} {\bf 23} (2018).

\bibitem{nagamochi94}
H.~Nagamochi, T.~Ono, T.~Ibaraki, Implementing an efficient minimum capacity
  cut algorithm.
\newblock {\it Math. Program.\/} {\bf 67}, 325–341 (1994).

\bibitem{Jakatdar2022}
A.~Jakatdar, B.~Liu, T.~Warnow, G.~Chacko, {AOC}: Assembling overlapping
  communities.
\newblock {\it Quantitative Science Studies\/} {\bf 3}, 1079--1096 (2022).

\bibitem{Leskovec2005}
J.~Leskovec, J.~Kleinberg, C.~Faloutsos, {\it Proceedings of the eleventh {ACM}
  {SIGKDD} international conference on Knowledge discovery in data mining\/}
  ({ACM}, 2005), pp. 177--187.

\bibitem{Yang2013}
J.~Yang, J.~Leskovec, Defining and evaluating network communities based on
  ground-truth.
\newblock {\it Knowledge and Information Systems\/} {\bf 42}, 181--213 (2013).

\bibitem{Leskovec2010}
J.~Leskovec, D.~Huttenlocher, J.~Kleinberg, {\it Proceedings of the {SIGCHI}
  Conference on Human Factors in Computing Systems\/} ({ACM}, 2010), pp.
  1361--1370.

\bibitem{Yin2017}
H.~Yin, A.~R. Benson, J.~Leskovec, D.~F. Gleich, {\it Proceedings of the 23rd
  {ACM} {SIGKDD} International Conference on Knowledge Discovery and Data
  Mining\/} ({ACM}, 2017), pp. 555--564.

\bibitem{albert2002statistical}
R.~Albert, A.-L. Barab{\'a}si, Statistical mechanics of complex networks.
\newblock {\it Reviews of modern physics\/} {\bf 74}, 47 (2002).

\bibitem{fortunato-resources}
S.~Fortunato, Resources (2023). {https://www.santofortunato.net/resources}.

\bibitem{hagberg2008exploring}
A.~Hagberg, P.~Swart, D.~S~Chult, Exploring network structure, dynamics, and
  function using networkx, {\it Tech. rep.\/}, Los Alamos National Lab.(LANL),
  Los Alamos, NM (United States) (2008).

\bibitem{lfr-generation-code}
Y.~Tabatabaee, Emulating real networks using {LFR} graphs,
  \url{https://github.com/ytabatabaee/emulate-real-nets} (2023).

\bibitem{clauset2009power}
A.~Clauset, C.~R. Shalizi, M.~E. Newman, Power-law distributions in empirical
  data.
\newblock {\it SIAM review\/} {\bf 51}, 661--703 (2009).

\bibitem{alstott2014powerlaw}
J.~Alstott, E.~Bullmore, D.~Plenz, powerlaw: a python package for analysis of
  heavy-tailed distributions.
\newblock {\it PloS one\/} {\bf 9}, e85777 (2014).

\bibitem{slota2019scalable}
G.~M. Slota, J.~W. Berry, S.~D. Hammond, S.~L. Olivier, C.~A. Phillips,
  S.~Rajamanickam, {\it Proceedings of the International Conference for High
  Performance Computing, Networking, Storage and Analysis\/} (2019), pp. 1--14.

\bibitem{pedregosa2011scikit}
F.~Pedregosa, G.~Varoquaux, A.~Gramfort, V.~Michel, B.~Thirion, O.~Grisel,
  M.~Blondel, P.~Prettenhofer, R.~Weiss, V.~Dubourg, {\it et~al.\/},
  Scikit-learn: Machine learning in python.
\newblock {\it the Journal of machine Learning research\/} {\bf 12}, 2825--2830
  (2011).

\bibitem{leiden-code}
V.~Traag, Leiden algorithm: leidenalg,
  \url{https://github.com/vtraag/leidenalg} (2019).

\bibitem{ikc-code}
E.~Wedell, M.~Park, Iterative k-core software,
  \url{https://github.com/chackoge/ERNIE_Plus/blob/master/Illinois/clustering/eleanor/code/IKC.py}
  (2021).

\end{thebibliography}
\bibliographystyle{ScienceAdvances}

\noindent \textbf{Acknowledgements:}  The authors thank Nathan Bryans, Christine Ballard, and Bryan Barker from Oracle Research for their assistance in setting up and using the Oracle Cloud Infrastructure,
\\
\noindent \textbf{Funding:} This work was supported in part by Insper-Illinois Collaboration and by a Research Award to TW from Oracle Research..\\
\noindent \textbf{Author Contributions} GC and TW conceived the research; GC, TW, and YT designed the analyses. 
VR, FA, MP, BL, YT, VP, DK, and RR developed software to support analyses.  
MP, YT, GC, VR, VP conducted the analyses. GC, TW, YT, and MP interpreted the data and wrote the manuscript.
\\
\noindent \textbf{Competing Interests} The authors declare that they have no competing financial interests.\\
\noindent \textbf{Data and materials availability:} Additional data and materials are available online.

\end{document}